\documentclass[preprint,9pt]{elsarticle}

\usepackage{amssymb}
\usepackage{amsmath}
\usepackage{xcolor}
\usepackage[colorlinks=true, linkcolor=blue, citecolor=blue, urlcolor=blue]{hyperref}
\usepackage[margin=1in]{geometry}     % For margin adjustments
\usepackage{bm}% bold math
\usepackage[ruled,vlined]{algorithm2e}
\usepackage{booktabs}

\usepackage{seqsplit}
\usepackage{xargs}
\usepackage{xstring}
\usepackage[colorinlistoftodos,prependcaption,textsize=small]{todonotes}

\newcommandx{\subitems}[2][1=]{}
\newcommandx{\subsubitems}[2][1=]{}
\newcommandx{\question}[2][1=]{}

\todostyle{ai}{inline, linecolor=teal,backgroundcolor=teal!15,bordercolor=teal}
\todostyle{qw}{inline, linecolor=blue,backgroundcolor=blue!15,bordercolor=blue}
\todostyle{ma}{inline, linecolor=orange,backgroundcolor=orange!15,bordercolor=orange}
\todostyle{xl}{inline, linecolor=gray,backgroundcolor=gray!15,bordercolor=gray}
\todostyle{unknown}{inline, linecolor=yellow,backgroundcolor=yellow!15,bordercolor=yellow}
\todostyle{done}{inline, linecolor=black,backgroundcolor=black!15,bordercolor=black}

\journal{Journal of Computational Physics}

\begin{document}

\begin{frontmatter}

\title{Pressure reconstruction from error-embedded gradient measurements: a Gaussian-process generalization of Green's function integration}

\author[sdsu]{Zejian You} %% Author name

\author[sdsu]{Mohamed Amine Abassi}
\author[sdsu]{Xiaofeng Liu}
\author[sdsu]{Qi Wang\corref{cor1}}
\ead{qwang4@sdsu.edu}

\cortext[cor1]{Corresponding author}

\affiliation[sdsu]{organization={San Diego State University},%Department and Organization
            addressline={5500 Campanile Drive}, 
            city={San Diego},
            postcode={92182}, 
            state={CA},
            country={USA}}

%% Abstract
\begin{abstract}
Reconstructing scalar fields from error-embedded gradient measurements is a fundamental linear inverse problem with broad applications in computational physics. Conventional approaches, such as Poisson-based solvers and the Green's Function Integration (GFI) method, require explicit boundary conditions extracted from the same error-embedded observations. In this study, we assess the accuracy of a Gaussian Process Regression (GPR) framework for reconstructing pressure fields in turbulent flows from error-embedded pressure-gradient data derived from kinematic measurements. The probabilistic nature of GPR inherently provides tunable denoising, eliminates the need for boundary conditions, and produces a pointwise posterior-variance error estimate.
A central theoretical result of the present work is that GFI is the noiseless limit of GPR, which on the unbounded plane reduces to the well-known logarithmic kernel and in three dimensions to the inverse-distance kernel. 
We verify this convergence numerically on a fine planar grid, where the singular spectrum of the GPR operator merges into that of GFI as the assumed observation noise parameter is reduced, with the two operators becoming indistinguishable until the high-wavenumber kernel cutoff.
The framework is validated on two-dimensional slices and three-dimensional subdomains of a forced homogeneous isotropic turbulence from the Johns Hopkins Turbulence Database. With an empirical mixture-of-Gaussians (MoG-$3$) kernel fitted directly to the pressure correlation function, GPR performs at least as well as GFI. In situations with under-resolved data or high noise, GPR outperforms GFI, while delivering a calibrated pointwise posterior uncertainty whose standardized residuals satisfy $|z|<2$ over $95\%$ of grid points. The framework extends to three dimensions through a tensor-product Kronecker solver coupled to conjugate gradients with close to $\mathcal{O}(N^3\log N)$ cost. A closed-form error lower bound on a periodic cube is derived for the GPR operator, with the residual gap attributable to boundary contamination on non-periodic finite domains.
\end{abstract}

% \includegraphics[width=\textwidth]{Figures/graphical_abstract.png}

% \begin{highlights}
% \item Gaussian Process Regression (GPR) reconstructs pressure from error-embedded gradient data without explicit boundary conditions and provides a calibrated pointwise posterior uncertainty.
% \item GPR generalizes the Green's Function Integration (GFI) method: GFI is recovered as the noiseless limit (i.e., the assumed observation noise parameter approaches zero) of GPR. This convergence is verified through SVD spectral analysis of both operators.
% \item An empirical mixture-of-Gaussians (MoG-$3$) kernel fitted directly to the pressure correlation function eliminates the parametric kernel-family choice and preserves a fast Kronecker structure for highly parallel evaluation with a computational complexity close to $\mathcal{O}(N^3\log(N))$ for three-dimensional reconstruction.
% \item On JHTDB isotropic turbulence in 2D and 3D, GPR matches GFI at low noise with fine resolution, and outperforms it at high noise or coarse resolution.
% \end{highlights}

% %% Keywords
% \begin{keyword}
% Gaussian Process Regression, Pressure Reconstruction, Uncertainty Quantification, Omni-Directional Integration

% %% PACS codes here, in the form: \PACS code \sep code

% %% MSC codes here, in the form: \MSC code \sep code
% %% or \MSC[2008] code \sep code (2000 is the default)
% \MSC[2008] 15-00 \sep 35R30 
% \end{keyword}

\end{frontmatter}



\section{Introduction}
Reconstructing a scalar field from error-embedded measurements of its gradient is a fundamental inverse problem in computational physics. In fluid mechanics, Particle Image Velocimetry (PIV) provides instantaneous velocity fields from which the pressure gradient can be obtained from the Navier--Stokes momentum equation, with the viscous contribution often negligible away from solid walls at sufficiently high Reynolds number \citep{liu2003measurements, liu2006instantaneous}. Accurate recovery of the instantaneous pressure field is essential for studying aeroacoustic noise generation \citep{buchta2017near, wu2008methods}, boundary-layer separation \citep{gramann1990detection, cerretelli2009boundary}, turbulence dynamics \citep{yeung2012dissipation}, and the reconstruction of pressure boundary data for Poisson-based Navier--Stokes solvers \citep{abassi2025adjoint}, while direct non-intrusive pressure measurements remain challenging in many experimental settings. 
More broadly, this task shares the central philosophy of data assimilation: incomplete and error-embedded observations are combined with physical constraints to infer an unobserved field of the system \citep{zaki2025turbulence}. In this sense, pressure reconstruction from PIV may be viewed as a spatial field-inference problem closely related to data assimilation, but focused on recovering an instantaneous pressure field from experimentally accessible kinematic information. 

Three families of methods address this reconstruction problem, and understanding their relationship is the starting point of the present work. Poisson-based solvers differentiate the gradient data to form a source term and invert the Laplacian globally, requiring explicit Neumann boundary conditions extracted from the same error-embedded observations \citep{wu2008methods}. The Omni-Directional Integration (ODI) method \citep{liu2003measurements, liu2006instantaneous, liu2016instantaneous, liu2018pressure, moreto2022experimentally}, instead of seeking solutions to Poisson Equation, reconstructs pressure distributions directly from its measured gradient. It utilizes the ``path independence'' property of scalar gradient line integration, and avoids the Neumann boundary condition step by averaging line integrals from all directions and obtaining the Dirichlet boundary condition. Summation of line integration along independent integration paths helps minimize or cancel errors and provide a denoising effect that often requires dense paths with appropriate density levels to achieve the optimal reconstruction result \citep{liu2020error}. Theoretically, in the continuous limit of infinitely dense ODI paths, averaging the line integration converges to a surface integration. An application of Green's identity shows that ODI converges to the Green's Function Integration (GFI) method under that limit \citep{wang2023green}, which is equivalent to convolution with the gradient of the Newtonian Green's function on the measurement domain. This is the same operator as solving the Poisson equation away from the boundary. At full resolution, ODI, GFI, and the Poisson solver are mathematically the same linear operator, and practical differences arise entirely from how the singular Newtonian Green's function is approximated on a finite grid and how the boundary condition is treated. Matrix formulations of ODI \citep{zigunov2024one} make this discrete structure explicit and recover the computational efficiency of Poisson solvers while retaining the ODI averaging structure.

This equivalence is followed by a gap in the literature: the denoising behavior of these methods is controlled entirely by the grid-level discretization of the singular Green's function, not by any prior knowledge of the pressure field. Every grid-based implementation implicitly replaces the singular kernel with a bandlimited approximation whose effective cutoff wavenumber is set by grid or ray spacing. The resulting low-pass filter attenuates measurement error, but its shape is determined by numerical practicality rather than by the spatial correlation structure of the pressure field, and changing the denoising strength requires changing the discretization. In principle, a denoising solver can be designed to fully incorporate these statistical structures of the scalar field, such as the characteristic correlation length and the algebraically decaying energy spectrum of turbulent pressure fields, into the reconstruction. The application of existing methods has been adopted in an empirical rather than principled manner.

Gaussian Process Regression (GPR) \citep{rasmussen2006gaussian, schulz2018tutorial, mons2019kriging, you2023pressure} provides a probabilistic framework for reconstructing latent fields from sparse, error-embedded, or indirect observations. In GPR, the pressure field is modeled as a Gaussian random field whose covariance kernel encodes prior assumptions on smoothness, correlation length, anisotropy, and spectral content. Conditioned on measurements, the posterior mean yields a regularized reconstruction, while the posterior covariance quantifies spatially varying uncertainty. This is particularly attractive for pressure reconstruction from PIV-derived quantities, where measurement error, incomplete spatial coverage, and modeling error are intrinsic. A key advantage in the present setting is that linear functionals of a Gaussian process remain Gaussian, so gradient and other derivative observations can be incorporated analytically through derivative-kernel cross-covariances \citep{solak2003derivative, rasmussen2006gaussian}. The denoising behavior is therefore governed explicitly by the assumed covariance model, rather than arising only as a by-product of numerical discretization.

The broader mathematical and methodological context is given by Reproducing Kernel Hilbert Space (RKHS) and kernel methods \citep{wahba1990spline, hastie2009elements, kimeldorf1971representer, kanagawa2018gaussian}. From this viewpoint, the GPR posterior mean is closely related to kernel-ridge or Tikhonov-regularized estimation \citep{smola1998learning}, while the covariance kernel specifies the function class in which smoothness and correlation assumptions are imposed. Building on this foundation, a growing literature has extended Gaussian-process methods to derivative-constrained, boundary-constrained, and PDE-informed inverse problems, including probabilistic solvers and physics-informed formulations that incorporate linear operators and error-embedded observations directly into the inference procedure \citep{graepel2003solving, solak2003derivative, cockayne2017probabilistic, raissi2017machine, swiler2020survey, valentine2020gaussian, dalton2024boundary, pfortner2022physics, li2025gaussian}. More broadly, related field-inversion paradigms infer spatially distributed latent correction fields from data rather than only a small set of parameters, providing a useful parallel development in computational physics and fluid mechanics \citep{parish2016paradigm, wang2019discrete}.

Throughout this paper, ``noise'' refers to experimental measurement error,  distinct from physically meaningful fluctuations such as acoustic noise. We derive the GPR formulation for pressure reconstruction from error-embedded gradient data, together with an explicit identification of GFI as the noiseless limit (i.e. the assumed observation noise parameter approaches to zero) of GPR through RKHS theory and a characterization of the denoising-versus-fidelity trade-off through singular value decomposition and impulse-response analyses. We use the empirical correlation function of the pressure field directly as the GPR prior covariance, via a positive-definite mixture-of-Gaussians fit, which eliminates the parametric kernel-family choice and approaches the linear-estimator Wiener bound on the noise-power gain. The extension of the framework to three dimensions is done through a tensor-product Kronecker solver coupled to conjugate gradients \citep{hestenes1952methods}, which reduces the per-reconstruction cost from a fully dense problem to one that scales as the square of the planar size times the iteration count. All methods are validated on forced homogeneous isotropic turbulence from the Johns Hopkins Turbulence Database (JHTDB) \citep{li2008public, perlman2007data}, where GPR with optimized hyperparameters matches GFI at low noise with a fine observation grid, and outperforms it at high noise or a coarse resolution.
The remainder of the paper is organized as follows. Section \ref{sec:formulation} develops the mathematical framework and draws the relation between GPR and GFI. Section \ref{sec:problem} presents the two-dimensional validation and analysis, including the spectral demonstration of GPR/GFI convergence and the uncertainty-quantification study. Section \ref{sec:3Dresults} covers the three-dimensional results, including the tensor-product solver and uncertainty quantification. Section \ref{sec:conclusion} presents the conclusions and outlines future work.

\section{Mathematical formulation}
\label{sec:formulation}
\subsection{The Gaussian Process Regression formulation}
\label{sec:GPR}
We formulate the reconstruction problem within the framework of \emph{Gaussian Process Regression} (GPR)~\cite[see][for a concise introduction]{schulz2018tutorial}. 
The unknown scalar field is modeled as a Gaussian process, and gradient measurements are incorporated through derivative observations of the covariance kernel. 
This formulation provides a consistent treatment of measurement noise and yields both a reconstructed field and an associated uncertainty estimate.

In GPR, the observed pressure gradient at a spatial location $\boldsymbol{x}$ is modeled as
\begin{equation}
\nabla p_{\mathrm{obs}}(\boldsymbol{x}) = \nabla \tilde{p}(\boldsymbol{x}) + \varepsilon,
\end{equation}
where $\nabla \tilde{p}(\boldsymbol{x})$ denotes the true pressure gradient, $\varepsilon$ represents additive Gaussian noise, $\varepsilon \sim \mathcal{N}(0, \sigma_\varepsilon^2)$, and the subscript ``obs'' marks the observed (error-embedded) gradient field; the symbol $p_{\mathrm{obs}}$ stands for any scalar field whose gradient matches this observation, and is used throughout as the input quantity supplied to the reconstruction.

The pressure field itself is treated as a realization of a Gaussian process in an infinite-dimensional function space,
\begin{equation}
p(\boldsymbol{x}) \sim \mathcal{GP}\!\left(\bar{p}(\boldsymbol{x}),\, \mathcal{C}(\boldsymbol{x}, \boldsymbol{x}^{\prime})\right),
\end{equation}
where $\bar{p}(\boldsymbol{x}) = \mathbb{E}[p(\boldsymbol{x})]$ is the mean (or prior expectation) of the pressure field, 
and $\mathcal{C}(\boldsymbol{x}, \boldsymbol{x}^{\prime})$ is the covariance kernel describing the spatial correlation between 
$p(\boldsymbol{x})$ and $p(\boldsymbol{x}^{\prime})$. 
Namely,
\begin{equation}
\mathcal{C}(\boldsymbol{x}, \boldsymbol{x}^{\prime}) = 
\mathbb{E}\!\left[(p(\boldsymbol{x}) - \bar{p}(\boldsymbol{x}))(p(\boldsymbol{x}^{\prime}) - \bar{p}(\boldsymbol{x}^{\prime}))\right].
\end{equation}

For a continuous and stationary random process, parametric \emph{radial basis function} (RBF) kernels such as the Gaussian are commonly employed; the choice of $\mathcal{C}$ used in this paper is deferred to \S\ref{sec:hyper}, where we calibrate it directly against the empirical correlation function of the test field. During inference, $\sigma(\boldsymbol{x})$ is updated from the data, providing a natural quantification of uncertainty in the reconstructed pressure field.

The \emph{correlation function} associated with the kernel is defined as
\begin{equation}
\mathcal{K}(\boldsymbol{x}, \boldsymbol{x}^{\prime}) =
\frac{\mathcal{C}(\boldsymbol{x}, \boldsymbol{x}^{\prime})}
{\sigma(\boldsymbol{x}) \sigma(\boldsymbol{x}^{\prime})},
\label{eq:correlation_function}
\end{equation}
where $\sigma$ denotes the root-mean-square of the entire pressure field. For any stationary RBF kernel, $\mathcal{K}(\boldsymbol{x}, \boldsymbol{x}^{\prime})$ depends only on the Euclidean distance $r = \|\boldsymbol{x} - \boldsymbol{x}^{\prime}\|$, i.e., $\mathcal{K}(\boldsymbol{x}, \boldsymbol{x}^{\prime}) = \mathcal{K}(r)$.

\begin{figure}
    \centering
    \includegraphics[width=1\textwidth]{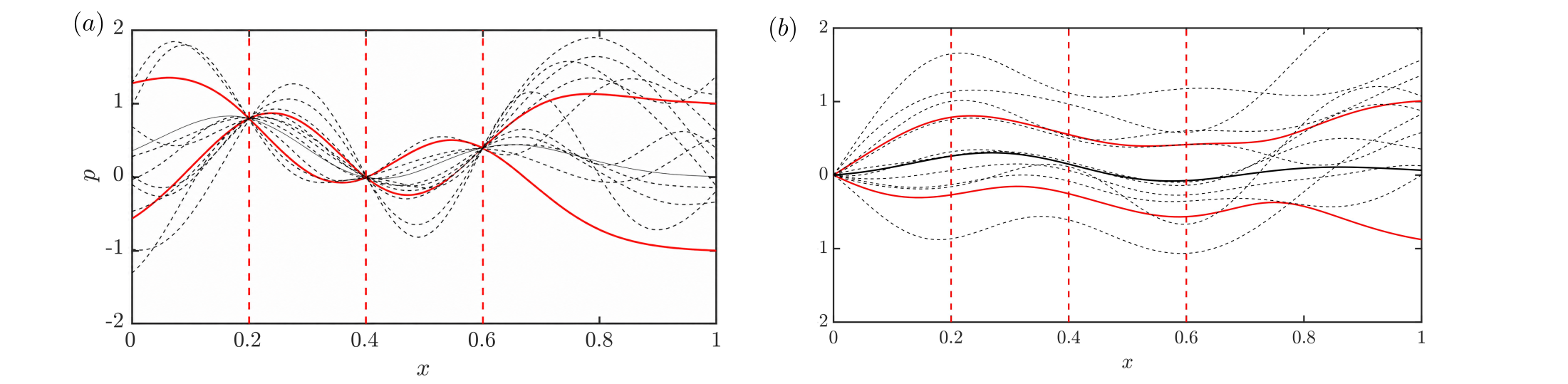}
    \caption{(a) An example of one-dimensional GPR when observations of the values of a smooth function are available. Red dashed lines mark the location of observations, while the black dashed lines are samples drawn from the posterior distribution. Two red solid lines mark the $\pm\sigma$ envelope of the posterior.
    (b) When observations of function derivatives are available, a similar approach can also be adopted to infer the values of the function, with an additional constraint that the function is zero at $x=0$.}
    \label{fig:1D_examples}
\end{figure}
    
While most applications of GPR are for observations of the function value directly \citep{mons2019kriging}, as shown in figure~\ref{fig:1D_examples}(a),
the formulation of GPR is general and can be applied to observations of the function gradient. An example of such reconstruction for a one-dimensional case is shown in Figure~\ref{fig:1D_examples}(b), where the dashed lines mark the observation locations for the gradient of the function $p(x)$.
Given $\boldsymbol{X}_*$, a vector of $n$ spatial locations $\boldsymbol{x_n}$ where data are observed, the samples of the pressure gradient observations form a multivariate Gaussian distribution,
\begin{equation}
    \boldsymbol{O} = \nabla p_{\mathrm{obs}}\!\left(\boldsymbol{X}_*\right) \sim \mathcal{N}\left( \nabla \bar{p}\big|_{\boldsymbol{X}_*}, \nabla_{\boldsymbol{x}}\nabla_{\boldsymbol{x}^{\prime}} \mathcal{C} \left(\boldsymbol{X}_*,\boldsymbol{X}_*\right) + \sigma_{\varepsilon}^2 \mathbf{I}_* \right).
\end{equation}
Here $\sigma_{\varepsilon}$ is the assumed noise level of synthetic noise introduced as a mimic of experimental data. $\mathbf{I}_*$ is the identity matrix. 
Once the observations are drawn from the above distribution, the observations of the pressure gradient at $\boldsymbol{X}_*$ and the unknown values of the pressure field at $\boldsymbol{X}$, a vector collection of spatial locations where reconstructions are conducted, follow the joint Gaussian distribution
\begin{equation}
    \begin{bmatrix}
        \boldsymbol{O}\\\\
        p(\boldsymbol{X})
    \end{bmatrix} \sim \mathcal{N}\left(
    \begin{bmatrix}
        \nabla \bar{p}\big|_{\boldsymbol{X}_*}\\\\
        \bar{p}(\boldsymbol{X})
    \end{bmatrix},
    \begin{bmatrix}
    \underbrace{\nabla_{\boldsymbol{x}}\nabla_{\boldsymbol{x}^{\prime}} \mathcal{C} \left(\boldsymbol{X}_*,\boldsymbol{X}_*\right) + \sigma_{\varepsilon}^2 \mathbf{I}_*}_{K_{gg}+\sigma_\varepsilon^2\mathbf{I}_*},  \underbrace{\nabla_{\boldsymbol{x}^{\prime}} \mathcal{C} \left(\boldsymbol{X},\boldsymbol{X}_*\right)}_{K_{pg}}\\
        \underbrace{\nabla_{\boldsymbol{x}} \mathcal{C} \left(\boldsymbol{X}_*,\boldsymbol{X}\right)}_{K_{gp}=K_{pg}^{\!\top}}, \underbrace{\mathcal{C} \left(\boldsymbol{X},\boldsymbol{X}\right)}_{K_{pp}}
    \end{bmatrix}
    \right).
\end{equation}
Throughout the paper, we use the uppercase symbol $K$ with subscripts to denote the discrete \emph{matrix} form of a kernel function evaluated at a finite set of grid points; the subscripts indicate the variates each block couples, with $p$ denoting the pressure and $g$ the pressure gradient. Specifically, $K_{pp}$ samples $\mathcal{C}$, the cross-covariance $K_{pg}$ samples $\nabla_{\!\boldsymbol{x}'}\mathcal{C}$, the gradient Gram $K_{gg}$ samples $\nabla_{\!\boldsymbol{x}}\nabla_{\!\boldsymbol{x}'}^{\!\top}\mathcal{C}$, with $K_{gp}=K_{pg}^{\!\top}$ by symmetry of $\mathcal{C}$.

The remaining piece of missing information is the reference pressure.
Since adding any constant value to the pressure field does not influence the dynamic structure of reconstructed pressure fields, we include one additional observation of the averaged pressure being zero within the domain in the formulation.
If the measurements of the pressure gradient are error-embedded, the pressure field can be recovered using Bayes' theorem, and the posterior conditional distribution is given by
\begin{equation}
    p\left(\boldsymbol{X}\right) \sim \mathcal{N}\!\left(\,\bar{p}+K_{pg}\bigl(K_{gg}+\sigma_\varepsilon^2\mathbf{I}_*\bigr)^{-1}\!\left(\boldsymbol{O}-\nabla\bar{p}\big|_{\boldsymbol{X}_*}\right),\; K_{pp}-K_{pg}\bigl(K_{gg}+\sigma_\varepsilon^2\mathbf{I}_*\bigr)^{-1}K_{gp}\right).
\end{equation}
The updated mean of the posterior distribution is regarded as the reconstructed pressure field from GPR, $p_{GPR}(\boldsymbol{X}) = \bar{p}+K_{pg}\bigl(K_{gg}+\sigma_\varepsilon^2\mathbf{I}_*\bigr)^{-1}\!\left(\boldsymbol{O}-\nabla\bar{p}\big|_{\boldsymbol{X}_*}\right)$.
Notice that the direct matrix inversion in the above expression requires $\mathcal{O}(N^{3d})$ operations, with $N$ being the numerical resolution, or the number of observations in each direction, and $d$ the dimensionality of the problem.
The direct inversion is therefore computationally intractable for large $N$ or $d$.
%Nevertheless, on structured Cartesian grids with a separable kernel, the Kronecker structure of the covariance blocks admits a matrix-free conjugate-gradient solve~\citep{hestenes1952methods} at cost $\mathcal{O}(N^{d+1}\, n_\mathrm{CG})$ per reconstruction rather than $\mathcal{O}(N^{3d})$, restoring tractability without sacrificing the full Bayesian posterior.
%This tensor-product formulation is developed in \S\ref{sec:tensor_gpr} and is the form actually used for all two- and three-dimensional reconstructions reported in \S\ref{sec:problem} and \S\ref{sec:3Dresults}.

Moreover, the covariance matrix of the posterior probability distribution can be computed from $C_{\mathrm{GPR}}(\boldsymbol{X}, \boldsymbol{X}) = K_{pp}-K_{pg}\bigl(K_{gg}+\sigma_\varepsilon^2\mathbf{I}_*\bigr)^{-1}K_{gp}$.
Square roots of the diagonal elements of the above matrix are the pointwise posterior standard deviation $\sigma_{\mathrm{post}}(\boldsymbol{X})$, representing the uncertainty of the reconstructed pressure field at different spatial locations.

\subsection{Connection to Green's function integration}
\label{sec:GPR_GFI_connection}
GPR has a deep structural relationship with the \emph{Green's Function Integration} (GFI) method \citep{wang2023green}, an alternative gradient-to-pressure reconstruction we use as a reference throughout the paper. We record the relationship here, both to introduce GFI for unfamiliar readers and to make the explicit link between GFI and GPR, with the latter being the regularizer of the former.

GFI reconstructs pressure by integrating the measured gradient against the gradient of the Green's function $G$ of $-\nabla^{2}$ on $\Omega$:
\begin{equation}
    p_{\mathrm{GFI}}(\boldsymbol{x}) \;=\; \int_\Omega \nabla_{\!\boldsymbol{x}'}G(\boldsymbol{x},\boldsymbol{x}') \cdot \nabla p_{\mathrm{obs}}(\boldsymbol{x}')\, \mathrm{d}\boldsymbol{x}' \;+\; \text{boundary terms},
    \label{eq:GFI}
\end{equation}
where $G(\boldsymbol{x},\boldsymbol{x}')$ satisfies $-\nabla^{2}G(\,\cdot\,,\boldsymbol{x}')=\delta(\boldsymbol{x}-\boldsymbol{x}')$ and $\nabla_{\!\boldsymbol{x}'}G$ is the vector integration kernel mapping the gradient field at $\boldsymbol{x}'$ to the pressure at $\boldsymbol{x}$. On the unbounded plane, the Green's function is the logarithmic kernel $G(\boldsymbol{x},\boldsymbol{x}')=-(2\pi)^{-1}\log\lvert\boldsymbol{x}-\boldsymbol{x}'\rvert$, and in three dimensions the Newtonian kernel $G(\boldsymbol{x},\boldsymbol{x}')=(4\pi\lvert\boldsymbol{x}-\boldsymbol{x}'\rvert)^{-1}$. 
The boundary terms require solving an extra system of equations from compatibility conditions. Interested readers are referred to \citep{wang2023green} for the full details.

The GPR posterior mean of \S\ref{sec:formulation} takes the analogous integral form
\begin{equation}
    p_{\mathrm{GPR}}(\boldsymbol{x}) \;-\; \bar{p}(\boldsymbol{x})
    \;=\; \Bigl[\,K_{pg}\bigl(K_{gg}+\sigma_\varepsilon^2\mathbf{I}\bigr)^{-1}\!\bigl(\nabla p_{\mathrm{obs}}-\nabla\bar{p}\bigr)\Bigr](\boldsymbol{x}),
    \label{eq:GPR_integral}
\end{equation}
where $K_{pg}$ and $K_{gg}$ are the same prior covariance blocks as in \S\ref{sec:formulation}, now read as integral operators on $\Omega$ with kernels $\nabla_{\!\boldsymbol{x}'}\mathcal{C}(\boldsymbol{x},\boldsymbol{x}')$ and $\nabla_{\!\boldsymbol{x}}\nabla_{\!\boldsymbol{x}'}^{\!\top}\mathcal{C}(\boldsymbol{x},\boldsymbol{x}')$. Comparing \eqref{eq:GFI} and \eqref{eq:GPR_integral}, both methods integrate the observed gradient against a vector kernel: GFI uses the bare gradient of the Green's function $\nabla_{\!\boldsymbol{x}'}G(\boldsymbol{x},\boldsymbol{x}')$, while GPR uses $K_{pg}$ pre-conditioned by $(K_{gg}+\sigma_\varepsilon^2\mathbf{I})^{-1}$, with $\sigma_\varepsilon^2\mathbf{I}$ acting as Tikhonov regularization. In the noiseless limit $\sigma_{\varepsilon} \rightarrow 0$, and in an infinite (or periodic) domain, GPR turns out to be \emph{kernel-independent}, and returns to GFI.

On an infinite or periodic domain, stationarity diagonalizes every covariance block in Fourier space. Let the GPR kernel $\mathcal{C}(\boldsymbol{x},\boldsymbol{x}')=\mathcal{C}(\boldsymbol{r})$, $\boldsymbol{r}=\boldsymbol{x}-\boldsymbol{x}'$, have non-negative spectral density $\widehat{\mathcal{C}}(\boldsymbol{k})$. Differentiating $\mathcal{C}(\boldsymbol{r})$ once for the cross-covariance $K_{pg}$ and twice for the gradient Gram $K_{gg}$, the Fourier symbols read
\begin{equation}
    \widehat{K_{pg}}(\boldsymbol{k})
    \;=\; -\mathrm{i}\,\boldsymbol{k}\,\widehat{\mathcal{C}}(\boldsymbol{k}),
    \qquad
    \widehat{K_{gg}}(\boldsymbol{k})
    \;=\; \boldsymbol{k}\boldsymbol{k}^{\!\top}\,\widehat{\mathcal{C}}(\boldsymbol{k}),
    \label{eq:gradient_symbols}
\end{equation}
with $\widehat{K_{gg}}$ rank-$1$ along the longitudinal direction $\boldsymbol{k}/\lvert\boldsymbol{k}\rvert$ and the transverse subspace regularised solely by the noise variance $\sigma_\varepsilon^{2}\mathbf{I}$. On an infinite or periodic domain the boundary term in \eqref{eq:GFI} vanishes, so the GFI reconstruction is the curl-free integral of the observed gradient: $\widehat{\nabla p_{\mathrm{obs}}}(\boldsymbol{k})=\mathrm{i}\boldsymbol{k}\,\widehat{p_{\mathrm{GFI}}}(\boldsymbol{k})$. Inverting $\widehat{K_{gg}}+\sigma_\varepsilon^{2}\mathbf{I}$ reduces the GPR posterior mean to the per-mode Wiener filter
\begin{equation}
    \widehat{p_{\mathrm{GPR}}}(\boldsymbol{k})
    \;=\; \frac{\lvert\boldsymbol{k}\rvert^{2}\,\widehat{\mathcal{C}}(\boldsymbol{k})}
                {\lvert\boldsymbol{k}\rvert^{2}\,\widehat{\mathcal{C}}(\boldsymbol{k}) \;+\; \sigma_\varepsilon^{2}}\,
            \widehat{p_{\mathrm{GFI}}}(\boldsymbol{k})
    \;\xrightarrow[\sigma_\varepsilon\to 0]{}\;\widehat{p_{\mathrm{GFI}}}(\boldsymbol{k}),
    \label{eq:gpr_to_gfi_limit}
\end{equation}
i.e.\ GPR is a Wiener-smoothed version of the GFI reconstruction. The spectrum $\widehat{\mathcal{C}}$ enters numerator and denominator identically and cancels as $\sigma_\varepsilon\to 0$: \emph{every} positive-definite stationary GPR kernel reproduces GFI in this noiseless limit. The kernel choice influences the results at finite $\sigma_\varepsilon$, not the $\sigma_\varepsilon\!=\!0$ limit.

On a bounded domain $\Omega$, the Fourier diagonalization breaks down, and the kernel-agnostic argument needs modification: a particular kernel must encode the same boundary information that GFI carries through its dipole term. That kernel is uniquely determined by Reproducing Kernel Hilbert Space (RKHS) theory~\citep{aronszajn1950theory, wahba1990spline}. GFI returns the minimum-Dirichlet-energy interpolant of the gradient data,
\begin{equation}
    p_{\mathrm{GFI}} \;=\; \arg\min_{q}\; \tfrac{1}{2}\!\int_\Omega \lvert\nabla q(\boldsymbol{x})\rvert^{2}\, \mathrm{d}\boldsymbol{x}, \quad \text{constrained by  }\;\nabla q\big|_{\boldsymbol{X}_*} = \nabla p_{\mathrm{obs}}.
    \label{eq:gfi_dirichlet}
\end{equation}
with the proof in \S\ref{app:gfi_dirichlet}; this is the same Dirichlet-energy variational principle that underlies the pressure projection step of incompressible flow~\citep{taha2023minimization}. By Aronszajn's theorem~\citep{hastie2009elements} the RKHS whose squared norm is the Dirichlet energy in \eqref{eq:gfi_dirichlet} has reproducing kernel $G$, so $\mathcal{C}=G$ is the unique GPR covariance whose noiseless posterior coincides with GFI on a bounded $\Omega$, boundary terms included.

This RKHS selection of $\mathcal{C}=G$ also clarifies how the two methods handle $p\rvert_{\partial\Omega}$. The free-space $G$ is not positive definite, so the formal prior $\mathcal{GP}(0,G)$ is not a proper Gaussian measure (positive definite with finite variance). Its nullspace consists of the harmonic functions, which are uniquely determined by their boundary values. The boundary pressure must therefore be recovered as a latent unknown through the second-kind boundary integral equation generated by the dipole term in \eqref{eq:GFI}, as is the case in GFI.

A bounded positive-definite GPR covariance $\mathcal{C}$ with $\mathcal{C}(\boldsymbol{x},\boldsymbol{x})=\sigma_p^{2}<\infty$ is by contrast a proper Gaussian measure that constrains $p$ at every point of $\overline{\Omega}$ with finite prior variance, including the boundary, eliminating the boundary integral equation entirely. The two formulations are unified as the same Bayesian estimator with different priors on $p\rvert_{\partial\Omega}$: a flat improper prior gives GFI, a proper bounded-variance prior gives GPR.

\subsection{Three-dimensional reconstruction}
\label{sec:3d_recon}

 A direct application of GPR, as shown in \S\ref{sec:GPR}, to three dimensional problems requires the direct inversion of a gradient-observation Gram matrix of size $\mathcal{O}(N^3\times N^3)$ on an $N^3$ grid, with cost $\mathcal{O}(N^9)$ and storage $\mathcal{O}(N^6)$; this becomes prohibitive for all but the smallest problems. We consider two strategies that retain tractability. The first, \S\ref{sec:3d_recon:planewise}, applies the two-dimensional formulation on orthogonal families of planes and reconciles the plane-wise reconstructions by a least-squares choice of the integration constants. The second, \S\ref{sec:tensor_gpr}, exploits the Kronecker structure that a separable kernel imposes on the covariance, yielding a single three-dimensional inference at a cost competitive with the plane-wise procedure. 
%The two are compared on a JHTDB subvolume in \S\ref{sec:3d_tensor_comparison}.

\subsubsection{Plane-wise reconstruction with least-squares integration constants}
\label{sec:3d_recon:planewise}

The simplest dimensional decomposition applies the general formulation in \S\ref{sec:GPR} on three orthogonal families of planes (slices of constant $x$, constant $y$, and constant $z$), reducing the three-dimensional inference to $3N$ independent two-dimensional solves at total asymptotic cost $\mathcal{O}(N^{7})$ with inversions of 2D dense Gram matrices. Because integrating a two-dimensional gradient field recovers pressure only up to an additive constant, the three resulting volumes disagree by a single offset per slice along each axis. The offsets are chosen to minimize the volume-wise pairwise mismatch between the three reconstructions subject to the gauge condition that each offset family has zero mean, which admits a closed-form solution in terms of planar averages of the inter-volume differences. The reconciled three-dimensional reconstruction is then taken as the arithmetic mean of the three offset-corrected volumes. The full derivation, the resulting closed-form expressions, and the corresponding algorithm are collected in \ref{appx:ls_3d}.

\subsubsection{Tensor-product Kronecker Gaussian Process Regression on structured grids}
\label{sec:tensor_gpr}

The plane-wise strategy of \S\ref{sec:3d_recon:planewise} has a structural cost, since each plane is treated as an independent inference problem, and the cross-plane noise correlation is discarded. 
The inter-plane consistency is enforced only through a single additive offset per plane. On a structured Cartesian grid, the full three-dimensional posterior can in fact be computed at a cost competitive with the plane-wise procedure, by exploiting the Kronecker structure that a separable kernel imparts on the covariance.

We adopt a stationary product correlation function
\begin{equation}
    \mathcal{K}(\boldsymbol{r},\boldsymbol{r}') \;=\; \mathcal{K}_x(x,x')\,\mathcal{K}_y(y,y')\,\mathcal{K}_z(z,z'),
    \label{eq:sep_kernel}
\end{equation}
in which the three-dimensional correlation factorises as a product of three one-dimensional correlations along each Cartesian axis; the full prior covariance follows from \eqref{eq:correlation_function} as $\mathcal{C}(\boldsymbol{r},\boldsymbol{r}')=\sigma_p^{2}\,\mathcal{K}(\boldsymbol{r},\boldsymbol{r}')$. Several common kernels admit this product form, e.g., the mixture-of-Gaussians kernel, and the tensor-product machinery below applies to any $\mathcal{K}$ of the separable form \eqref{eq:sep_kernel}.

Enumerating the $N_xN_yN_z$ grid points in a fixed multi-index order $(i,j,k)\mapsto i+N_x(j+N_yk)$ and applying \eqref{eq:sep_kernel} to every pair of grid points gives, blockwise, the prior covariance matrix as a Kronecker product. Recall that for two matrices $A\in\mathbb{R}^{m\times n}$ and $B\in\mathbb{R}^{p\times q}$, the Kronecker product $A\otimes B\in\mathbb{R}^{(mp)\times(nq)}$ is the block matrix
\begin{equation}
    A\otimes B \;=\;
    \begin{pmatrix}
    A_{11}B & A_{12}B & \cdots & A_{1n}B \\
    A_{21}B & A_{22}B & \cdots & A_{2n}B \\
    \vdots  & \vdots  & \ddots & \vdots  \\
    A_{m1}B & A_{m2}B & \cdots & A_{mn}B
    \end{pmatrix},
    \qquad
    [A\otimes B]_{(i-1)p+k,\,(j-1)q+l} \;=\; A_{ij}\,B_{kl},
    \label{eq:kron_def}
\end{equation}
i.e.\ each scalar entry $A_{ij}$ is replaced by the scaled block $A_{ij}B$. The triple Kronecker product $A\otimes B\otimes C$ is defined iteratively as $(A\otimes B)\otimes C$, and the mixed-product rule $(A\otimes B)(\boldsymbol{v}\otimes\boldsymbol{w})=(A\boldsymbol{v})\otimes(B\boldsymbol{w})$ extends accordingly; this rule is what makes the matvec cheap and underlies the mode-product form developed below in \eqref{eq:mode_product}. With this definition, the prior covariance reads
\begin{equation}
    K_{pp} \;=\; \sigma_p^2\, K_x\otimes K_y\otimes K_z,
    \qquad [K_\alpha]_{ab} \;=\; \mathcal{K}_\alpha(x^{(\alpha)}_a,\, x^{(\alpha)}_b)\in\mathbb{R}^{N_\alpha\times N_\alpha}.
    \label{eq:Kpp_kron}
\end{equation}
Here the 1D matrix factors $K_x,K_y,K_z$ are the matrix forms of the 1D correlation factors $\mathcal{K}_x,\mathcal{K}_y,\mathcal{K}_z$ of the separable kernel.
Only the three one-dimensional factors $K_x,K_y,K_z$ of \eqref{eq:Kpp_kron} are ever stored; the full $(N_xN_yN_z)\times(N_xN_yN_z)$ matrix $K_{pp}$ is never formed.

Differentiating the 1D correlation $\mathcal{K}_\alpha(a,a')$ with respect to its first argument, its second argument, and both introduces three other matrices accompanying $K_{\alpha}$,
\begin{equation}
    [D_\alpha^{(1)}]_{ab}
    \!=\!\tfrac{\partial \mathcal{K}_\alpha(a,b)}{\partial a},
    \quad
    [D_\alpha^{(2)}]_{ab}
    \!=\!\tfrac{\partial \mathcal{K}_\alpha(a,b)}{\partial b},
    \quad
    [D_\alpha^{(11)}]_{ab}
    \!=\!\tfrac{\partial^2 \mathcal{K}_\alpha(a,b)}{\partial a\,\partial b}\,.
    \label{eq:D_building_blocks}
\end{equation}
By construction $D_\alpha^{(2)} = (D_\alpha^{(1)})^{\!\top}$. Every covariance block entering the three-dimensional system is built from \eqref{eq:D_building_blocks} by selecting, along each axis, whether a derivative acts on the first argument, the second, both, or neither.

Since each component $\partial_\alpha p$ ($\alpha\in\{x,y,z\}$) is a linear functional of a Gaussian process, the gradient $(\partial_x p,\partial_y p,\partial_z p)$ is jointly Gaussian with $p$ and the gradient Gram matrix decomposes into a $3\times 3$ block matrix with each block of size $(N_xN_yN_z)\times(N_xN_yN_z)$, each a triple Kronecker product of 1D factors selected according to which argument of which coordinate is differentiated:
\begin{equation}
    K_{gg}\!=\!\sigma_p^{2}\!
    \begin{bmatrix}
      D_x^{(11)}\!\otimes K_y\!\otimes K_z
      & D_x^{(1)}\!\otimes D_y^{(2)}\!\otimes K_z
      & D_x^{(1)}\!\otimes K_y\!\otimes D_z^{(2)} \\[4pt]
      D_x^{(2)}\!\otimes D_y^{(1)}\!\otimes K_z
      & K_x\!\otimes D_y^{(11)}\!\otimes K_z
      & K_x\!\otimes D_y^{(1)}\!\otimes D_z^{(2)} \\[4pt]
      D_x^{(2)}\!\otimes K_y\!\otimes D_z^{(1)}
      & K_x\!\otimes D_y^{(2)}\!\otimes D_z^{(1)}
      & K_x\!\otimes K_y\!\otimes D_z^{(11)}
    \end{bmatrix}.
    \label{eq:Kff_3x3}
\end{equation}
Symmetry $K_{gg}=K_{gg}^{\top}$ follows from $D_\alpha^{(2)}=(D_\alpha^{(1)})^{\!\top}$ applied entrywise. The prior-observation cross-covariance is a $1\times 3$ block row,
\begin{equation}
    K_{pg} \;=\; \sigma_p^{2}
    \begin{bmatrix}
      D_x^{(2)}\!\otimes K_y\!\otimes K_z
      & K_x\!\otimes D_y^{(2)}\!\otimes K_z
      & K_x\!\otimes K_y\!\otimes D_z^{(2)}
    \end{bmatrix}.
    \label{eq:Kpf_3}
\end{equation}
The observation-noise floor contributes $\sigma_\varepsilon^2 I$ to the $K_{gg}$ system, which is a Kronecker-compatible scaled identity.

A single block of \eqref{eq:Kff_3x3} has the generic form $M_x\otimes M_y\otimes M_z$, where $M_\alpha$ is a placeholder for whichever of $\{K_\alpha, D_\alpha^{(1)}, D_\alpha^{(2)}, D_\alpha^{(11)}\}$ that block selects. Its action on a grid tensor $V\in\mathbb{R}^{N_x\times N_y\times N_z}$ is three sequential mode products,
\begin{equation}
    \bigl[V\times_1 M_x \times_2 M_y \times_3 M_z\bigr]_{ijk}
    \;=\;
    \sum_{a,b,c} [M_x]_{ia}[M_y]_{jb}[M_z]_{kc}\, V_{abc},
    \label{eq:mode_product}
\end{equation}
implemented as three successive mode-$\alpha$ tensor--matrix contractions. The mode-$\alpha$ contraction (with $\alpha\in\{x,y,z\}$) acts on every $N_\alpha$-vector that lies along the $\alpha$-axis of $V$, and there are $N_xN_yN_z/N_\alpha$ such vectors. On the uniform Cartesian grid we use throughout this paper, each 1D factor $M_\alpha$ is Toeplitz (since the kernel is stationary) and admits an $\mathcal{O}(N_\alpha\log N_\alpha)$ FFT-based matvec via the standard $2N_\alpha$-circulant embedding~\citep{strang1986toeplitz} --- which gives an exact Toeplitz matvec on the original \emph{non-periodic} cube; the FFT only diagonalises the enlarged circulant operator, not the boundary condition. The mode-$\alpha$ contraction therefore costs $\mathcal{O}(N_xN_yN_z\,\log N_\alpha)$, and the three successive contractions cost $\mathcal{O}\!\bigl(N_xN_yN_z\sum_\alpha\log N_\alpha\bigr)$ per single Kronecker matvec. A matvec against the full block matrix in \eqref{eq:Kff_3x3} costs $\mathcal{O}\!\bigl(9\,N_xN_yN_z\sum_\alpha\log N_\alpha\bigr)$ instead of the $\mathcal{O}((3N_xN_yN_z)^{2})$ required for a dense application, while using $\mathcal{O}(\sum_\alpha N_\alpha^{2})$ storage for the 1D Toeplitz factors instead of $\mathcal{O}((3N_xN_yN_z)^{2})$ for the full Gram. On a cubic grid $N_x=N_y=N_z=N$ this reduces to $\mathcal{O}(N^{d}\log N)$ per matvec versus $\mathcal{O}(N^{2d})$ dense, with $\mathcal{O}(N^{2})$ storage for the 1D kernel factors $K_x,K_y,K_z$ versus $\mathcal{O}(N^{2d})$ for the full Gram. The total working memory of the algorithm is then $\mathcal{O}(N^{d})$, set by the gradient observations $\boldsymbol{O}$ and the CG iterates rather than by the kernel factors.

For scattered observation nodes the Toeplitz structure is lost, but the Fast Gauss Transform~\citep{greengard1991FGT} still delivers $\mathcal{O}(N\log(1/\epsilon))$ per 1D matvec for Gaussian kernels, preserving the $\mathcal{O}(N^{d}\log N)$ scaling.

The GPR posterior mean on the grid is
\begin{equation}
    p_{GPR} \;=\; K_{pg}\bigl(K_{gg}+\sigma_\varepsilon^{2}I\bigr)^{-1}\boldsymbol{O},
    \qquad \boldsymbol{O} = [\boldsymbol{O}_x;\,\boldsymbol{O}_y;\,\boldsymbol{O}_z],
    \label{eq:posterior_mean_3d}
\end{equation}
and is evaluated by the method of conjugate gradients (CG) \citep{hestenes1952methods}, an iterative Krylov solver for symmetric positive-definite linear systems that requires only matrix-vector products with the system operator and converges to machine precision in at most $r$ iterations, where $r$ is the number of distinct eigenvalues of the operator (in practice, far fewer iterations suffice for clustered spectra). Applied to the symmetric positive-definite system in \eqref{eq:posterior_mean_3d}, each CG iteration performs nine triple-Kronecker matvecs of the form \eqref{eq:mode_product}, one per block of \eqref{eq:Kff_3x3}, plus $\mathcal{O}(N_xN_yN_z)$ axpy operations; convergence is reached in tens of iterations for the turbulence data used in this work. 

As $\sigma_\varepsilon\to 0$, $K_{gg}$ develops a one-dimensional null space along the constant-$p$ direction, because gradient observations determine $p$ only up to an additive constant, and the linear system in \eqref{eq:posterior_mean_3d} becomes singular. In practice this is harmless, since even nominally noise-free data is regularized by a small jitter $\sigma_\varepsilon^{2}\!\sim\!10^{-12}\sigma_p^{2}$ that keeps $K_{gg}+\sigma_\varepsilon^{2}I$ uniformly positive definite without affecting the reconstruction. The integration-constant degeneracy that the gauge condition $\sum_{ijk}p_{ijk}=0$ resolves explicitly in the noiseless case is, in the error-embedded formulation, automatically pinned by the prior mean $\mathbb{E}[p]=0$ that enters through the term $\sigma_\varepsilon^{2}I$ in $K_{gg}$. The complete procedure is summarized in Algorithm~\ref{alg:tensor_gpr}.

\begin{algorithm}[tb]
\caption{Tensor-product 3D GPR\label{alg:tensor_gpr}}
\KwIn{discretised gradient observations $\boldsymbol{O}_x,\boldsymbol{O}_y,\boldsymbol{O}_z$ on an $N_x\!\times\!N_y\!\times\!N_z$ grid (each component the corresponding $\partial p/\partial x_\alpha$ at every grid point); 1D kernel factors $K_\alpha, D_\alpha^{(1)}, D_\alpha^{(2)}, D_\alpha^{(11)}$; hyperparameters $\sigma_p, \sigma_\varepsilon$ with $\sigma_\varepsilon>0$; Krylov tolerance $\tau$ (set to $10^{-8}$ throughout this paper)}
\KwOut{posterior-mean reconstruction $p_{GPR}\in\mathbb{R}^{N_x\times N_y\times N_z}$}
\SetKwFunction{TripleKron}{TripleKronProd}
\SetKwProg{Fn}{Function}{:}{end}
\Fn{\TripleKron{$V; M_x, M_y, M_z$}}{
  \tcp*[h]{$V\in\mathbb{R}^{N_x\times N_y\times N_z}$; $M_\alpha\in\mathbb{R}^{N_\alpha\times N_\alpha}$; returns $V\!\times_1\!M_x\!\times_2\!M_y\!\times_3\!M_z$, see \eqref{eq:mode_product}}\;
  \tcp{Mode-1: contract along $a$.}
  \For{$j=1,\dots,N_y$, $k=1,\dots,N_z$}{
    $W^{(1)}_{:,j,k} \leftarrow M_x\,V_{:,j,k}$ \tcp*{$N_x\!\times\!N_x$ Toeplitz matrix times $N_x$-vector}
  }
  \tcp{Mode-2: contract along $b$.}
  \For{$i=1,\dots,N_x$, $k=1,\dots,N_z$}{
    $W^{(2)}_{i,:,k} \leftarrow M_y\,W^{(1)}_{i,:,k}$ \tcp*{$N_y\!\times\!N_y$ Toeplitz matrix times $N_y$-vector}
  }
  \tcp{Mode-3: contract along $c$.}
  \For{$i=1,\dots,N_x$, $j=1,\dots,N_y$}{
    $W^{(3)}_{i,j,:} \leftarrow M_z\,W^{(2)}_{i,j,:}$ \tcp*{$N_z\!\times\!N_z$ Toeplitz matrix times $N_z$-vector}
  }
  \KwRet $W^{(3)}$\;
}
\BlankLine
Assemble the nine triple-Kronecker blocks of $K_{gg}$ in \eqref{eq:Kff_3x3} and the three of $K_{pg}$ in \eqref{eq:Kpf_3} as matrix-free objects: each block stores its three 1D factors $\{M_x,M_y,M_z\}$ and applies via \TripleKron{$\cdot;M_x,M_y,M_z$}. A matvec against the full $K_{gg}$ is the sum of nine such block calls; against $K_{pg}$, the sum of three\;
$\boldsymbol\beta \leftarrow \mathrm{CG}\bigl(K_{gg}+\sigma_\varepsilon^{2}I,\,\boldsymbol{O};\,\tau\bigr)$ \tcp*{symmetric positive-definite, since $\sigma_\varepsilon>0$}
$p_{GPR} \leftarrow K_{pg}\,\boldsymbol\beta$\;
\Return $p_{GPR}$
\end{algorithm}

\begin{table}[h!]
\centering
\label{tab:complexity_3d}
\begin{tabular}{lll}
\toprule
Formulation & Storage & Per-solve cost \\
\midrule
3D ODI\textsuperscript{$\dagger$} \citep{wang2019gpu}         & $\mathcal{O}(N^3)$    & $\mathcal{O}(N^5)$ \\
3D GFI  \citep{wang2023green}     & $\mathcal{O}(N^3)$    & $\mathcal{O}(N^6\cdot n_\mathrm{CG})$ \\
FFT 3D Poisson (periodic BCs)      & $\mathcal{O}(N^3)$  & $\mathcal{O}(N^3\log N)$ \\
3D multigrid Poisson (with BCs)        & $\mathcal{O}(N^3)$  & $\mathcal{O}(N^3 \cdot n_\mathrm{MG})$ \\
Dense 3D GPR (direct)        & $\mathcal{O}(N^6)$    & $\mathcal{O}(N^9)$ \\
Plane-wise 2D + LS           & $\mathcal{O}(N^4)$    & $\mathcal{O}(N^7)$ \\
Tensor-product 3D GPR & $\mathcal{O}(N^3)$    & $\mathcal{O}(N^3 \log(N) \cdot n_\mathrm{CG})$ \\
\bottomrule
\end{tabular}
\caption{Per-reconstruction complexity on a cubic $N\times N\times N$ grid. The CG iteration count $n_\mathrm{CG}$ for tensor-product GPR is independent of $N$ for smooth fields and is typically $\mathcal{O}(10^2)$. The multigrid V-cycle count $n_\mathrm{MG}$ is also independent of $N$ and is typically $\mathcal{O}(10)$. The two Poisson solvers are listed for reference and require explicit boundary data, which the gradient-only methods (ODI, GFI, plane-wise+LS, dense, and tensor-product GPR) do not. \textsuperscript{$\dagger$}The ODI complexities in this table are our own estimates from the parallel-ray algorithmic structure described in \citet{wang2019gpu} ($\mathcal{O}(N^{2})$ angular directions $\times\,\mathcal{O}(N^{2})$ rays per direction $\times\,\mathcal{O}(N)$ work per ray, plus an $\mathcal{O}(N^{3})$ accumulator on the grid); they are not stated explicitly in the cited reference.}
\end{table}
On a cubic $N\times N\times N$ grid, Table~\ref{tab:complexity_3d} summarises the per-reconstruction cost of the formulations considered in this paper alongside two boundary-value Poisson solvers (FFT-based with periodic boundary conditions, and geometric multigrid with prescribed Dirichlet or Neumann boundary conditions) included for reference. The boundary-value Poisson solvers are by far the cheapest at $\mathcal{O}(N^3 \log N)$ for FFT and $\mathcal{O}(N^3 \cdot n_\mathrm{MG})$ for multigrid (with $n_\mathrm{MG}$ a fixed number of V-cycles independent of $N$), but they require explicit boundary data and provide no built-in noise treatment. The tensor-product GPR recovers the full three-dimensional Bayesian posterior with an affordable asymptotic cost that scales with $N^3\log(N)$. 
The GPR algorithm using tensor products directly maps onto level-3 BLAS calls and is highly parallelizable on GPUs, so the wall time of a tensor-product solve falls roughly linearly with stream-multiprocessor count over the relevant range; this is a practical advantage that the dense-direct, ODI, and GFI baselines do not share.

\section{Gaussian process regression for two-dimensional pressure reconstruction}
\label{sec:problem}
% \begin{figure}
% \centering
% \begin{minipage}{.45\linewidth}
%     \includegraphics[width=\textwidth]{Figures/problem-pressure.png}
% \end{minipage}\quad%
% \begin{minipage}{.45\linewidth}
%     \includegraphics[width=\textwidth]{Figures/problem-dp.png}
% \end{minipage}%
% \caption{(a): True pressure field from an isotropic turbulence DNS database. (b)-(c): Pressure gradient obtained from the DNS pressure field by central finite difference method. (d)-(e): Sample realization of 1000 error embedded pressure gradient.}
% \label{fig:problem}
% \end{figure}
We test the GPR framework on a homogeneous isotropic turbulence flow field with Reynolds number $R_\lambda \sim 433$ based on Taylor microscale, using direct numerical simulation (DNS) data from the Johns Hopkins Turbulence Database (JHTDB) \citep{li2008public, perlman2007data, yeung2012dissipation} as a surrogate for an experimental measurement. The simulated data provides access to the fully resolved velocity and pressure fields, enabling controlled-noise tests of the reconstruction algorithm. 
Although the database is three-dimensional, both GFI and GPR reconstruct pressure on a two-dimensional plane given observations of the in-plane components of the gradient, so the same JHTDB realization supplies both the test problem and the ground-truth reference for the 2D benchmarks of this section.
The observations are the in-plane components of the pressure gradient sampled on an $x-y$ slice of the JHTDB \texttt{isotropic1024coarse} dataset.
The general numerical setup for \S\ref{sec:problem} is summarized in Table~\ref{tab:2d_setup}.

\begin{table}[h!]
\centering
\begin{tabular}{ll}
\toprule
dataset                                & JHTDB \texttt{isotropic1024coarse}, an $x-y$ slice \\
grid resolution                        & $256\times256$ \\
grid spacing $\Delta x$                & $2\pi/1024$ \\
domain                                 & $[0,\pi/2]\times[0,\pi/2]$ \\
prior amplitude $\sigma_p$             & $\mathrm{std}(p)\approx 0.362$ \\
noise model                            & additive Gaussian, $\boldsymbol{\eta}\sim\mathcal{N}(0,\sigma_\varepsilon^{2}I)$ with $\sigma_\varepsilon = \eta\,\max|\nabla p|$ \\
uniform-noise equivalent  & half-amplitude $\Delta = \sqrt{3}\,\sigma_\varepsilon$ on $[-\Delta,\Delta]$, same $\sigma_\varepsilon$ \\
\bottomrule
\end{tabular}
\caption{Numerical setup for the 2D benchmarks of \S\ref{sec:problem}.}
\label{tab:2d_setup}
\end{table}

We compare the proposed GPR reconstruction methods with GFI \citep{wang2023green}, as briefly introduced and compared in \S\ref{sec:GPR_GFI_connection}. 
The accuracy of the GPR posterior mean is governed almost entirely by the prior covariance kernel $\mathcal{C}(r)$. Two families of choices are common. The first is parametric: a one- or two-parameter family such as the Gaussian or Mat\'ern-$3/2$ kernel, whose length scale is tuned by cross-validation. The second is data-driven: take the empirical correlation function of the target field directly as the prior, removing the parametric kernel-family choice and approaching the linear-estimator Wiener bound on the noise-power gain. We adopt the data-driven route throughout this paper.

The empirical correlation function is computed by azimuthal averaging of the autocovariance,
\begin{equation}
    \mathcal{K}(\boldsymbol r) \;=\; \mathbb{E}\bigl[p(\boldsymbol x)\,p(\boldsymbol x + \boldsymbol r)\bigr],
    \qquad
    \mathcal{K}(r) \;=\; \frac{1}{2\pi}\int_0^{2\pi}\mathcal{K}\bigl(r\,\hat{\boldsymbol e}_\theta\bigr)\,d\theta,
    \quad \hat{\boldsymbol e}_\theta = (\cos\theta,\sin\theta),
    \label{eq:empirical_corr}
\end{equation}
with the expectation approximated by a spatial average over the JHTDB realization that serves as the ground-truth reference. By construction, $\mathcal{K}(r)$ has the spatial scales of the actual pressure field, so a GPR posterior mean built from $\mathcal{K}$ is automatically tuned to the spectrum of the data.

A complication arises from the positive-definiteness requirement. By Bochner's theorem \citep{bochner2005harmonic,stein1999interpolation}, a radial function $\mathcal{K}(r)$ qualifies as the covariance of an isotropic 2D Gaussian process if and only if its radial spectrum --- the zeroth-order Hankel transform of $\mathcal{K}$,
\begin{equation}
    S(|\boldsymbol{k}|) \;=\; \bigl(\mathcal{H}\mathcal{K}\bigr)\!\bigl(|\boldsymbol{k}|\bigr)
    \;\equiv\; 2\pi \int_0^\infty J_0(|\boldsymbol{k}|\,r)\,\mathcal{K}(r)\,r\,dr,
    \label{eq:bochner_spectrum}
\end{equation}
is everywhere non-negative. The 2D Fourier transform of a radial function reduces to this transform, so we use $\mathcal{H}$ rather than $\mathcal{F}$ throughout. The empirical correlation we compute on JHTDB has a pronounced negative lobe of depth $\approx -0.2$ for $r\in[0.5, 0.8]$ followed by a partial recovery toward $r\approx 1$. This anti-correlation ring is a real physical signature of the pressure field around coherent vortical structures, but results in a negative $S$ and a $\mathcal{K}(r)$ that is not positive-definite as a covariance kernel and cannot be used directly. The GPR prior must therefore sacrifice some features of $\mathcal{K}(r)$. The candidate positive-definition \emph{projections} of the correlation, the trade-offs they imply, and the resulting reconstruction errors are compared in \S\ref{sec:hyper}. The overall settlement is a positive-weight mixture of three Gaussians,
\begin{equation}
    \mathcal{K}_{\mathrm{emp}}(r) \;=\; \sum_{i=1}^{3}\alpha_i \exp\!\Bigl(-\frac{r^{2}}{2\,\ell_i^{2}}\Bigr),
    \qquad
    \sum_i \alpha_i = 1,
    \label{eq:mog_kernel}
\end{equation}
referred to throughout as the MoG-$3$ kernel. Each component is separable along Cartesian axes, so the Kronecker structure of the gradient Gram matrix from \S\ref{sec:formulation} carries over to MoG-$3$.

\begin{figure}[tb]
    \centering
    \includegraphics[width=\textwidth]{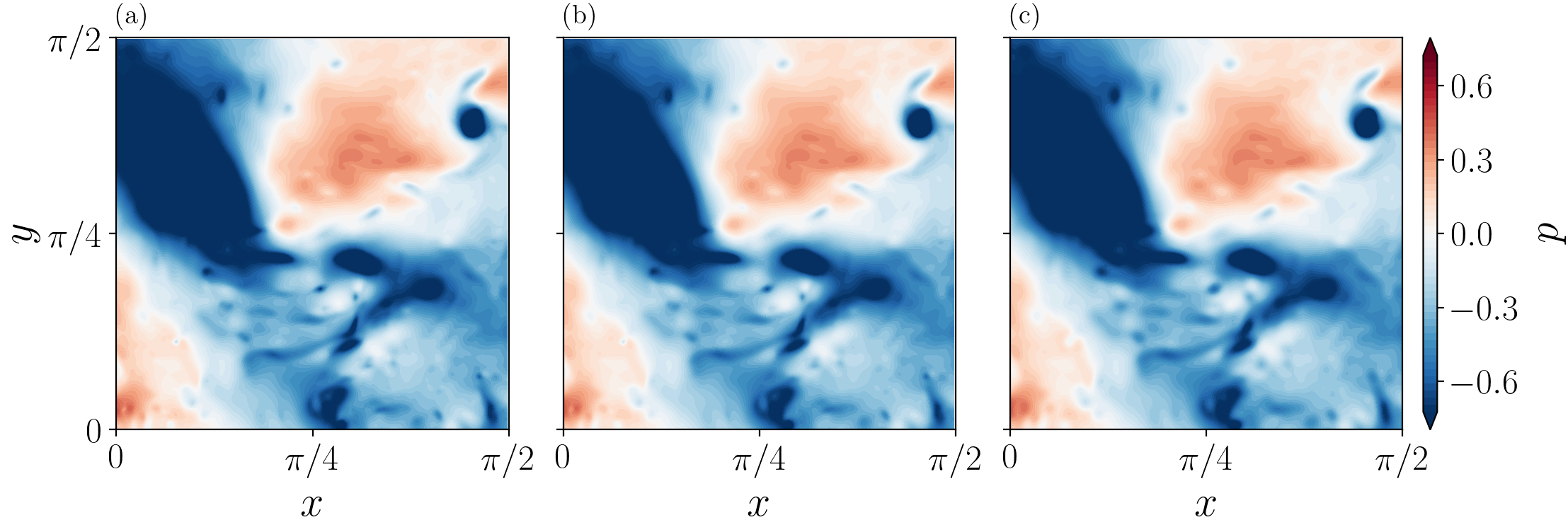}
    \caption{Noise-free reconstruction of the $256\times256$ JHTDB isotropic-turbulence pressure slice: (left) true field $p$; (center) GFI reconstruction ($\varepsilon_{\mathrm{GFI}}=0.008$); (right) full 2D GPR reconstruction using the empirical MoG-$3$ kernel \eqref{eq:mog_kernel} fitted to the true-field correlation function, with a tiny positive regularizer $\sigma_\varepsilon=10^{-2}\max|\nabla p|$ ($\varepsilon_{\mathrm{GPR}}=0.019$). Both methods recover the field to visual accuracy.}
    \label{fig:True_GPR_GFI}
\end{figure}

Figure~\ref{fig:True_GPR_GFI} serves as a noise-free verification of both reconstructions on the present 2D setup. With exact gradient observations and a small positive regularizer $\sigma_\varepsilon=10^{-2}\max|\nabla p|$ used in lieu of an exact zero in order to keep the Gram matrix strictly positive-definite, the empirical-kernel GPR attains relative RMSE $\varepsilon=0.020$ against the GFI baseline $\varepsilon=0.008$. The residual GPR error is governed by the prior bandwidth: the MoG-$3$ spectrum decays faster than the JHTDB pressure spectrum above $|\boldsymbol{k}|\sim 100$, so GPR attenuates fine-scale modes that the GFI baseline captures at its discretization floor.

\subsection{Empirical kernels: positive-definite projections and comparison}\label{sec:hyper}

We now examine three principled projections of the empirical correlation $\mathcal{K}(r)$ onto the cone of valid isotropic covariances and benchmark their reconstruction accuracy on the 2D setup of \S\ref{sec:problem}. Each projection preserves a different feature of the empirical curve, so the comparison quantifies what trade-off matters most for the reconstruction. For reference, we also include the two parametric kernel families (Gaussian and Mat\'ern-$3/2$) tuned to a single length scale by least-squares fit to the same empirical curve.

\begin{itemize}
  \item \textbf{MoG-$3$} (mixture of three Gaussians): a positive-weight sum of three Gaussians, fitted by non-linear least squares on the positive branch of $\mathcal{K}(r)$. Preserves the near-origin shape, discards the negative lobe. To preserve the Kronecker matvec structure of the gradient-observation solver (\S\ref{sec:formulation}) the prior covariance is given by \eqref{eq:mog_kernel}.
  \item \textbf{Spectral-clip}: $\mathcal{K}_{\mathrm{pd}}=\mathcal{H}^{-1}\!\max(\mathcal{H}\mathcal{K},\,0)$, i.e., clip the Hankel transform \eqref{eq:bochner_spectrum} of $\mathcal{K}$ to its non-negative part and invert. This is the $L^{2}$-nearest valid isotropic covariance (equivalently the Frobenius-nearest PSD matrix on a uniform grid).
  \item \textbf{Bessel-modulated Gaussian}: $\mathcal{K}_{\mathrm{pd}}(r)=a_1 e^{-r^{2}/2\ell_1^{2}}+a_2 J_0(k\,r)\,e^{-r^{2}/2\ell_2^{2}}$, $a_i\ge 0$, $\sum a_i = 1$, fitted by differential evolution with a penalty on the negative part of $S$. The oscillatory Bessel component is included specifically to reproduce a negative lobe.
\end{itemize}

Of these candidates, only MoG-$3$ is separable along Cartesian axes; the spectral-clip and Bessel-modulated Gaussian projections do not factor across coordinates and therefore destroy the Kronecker structure that the tensor-product solver of \S\ref{sec:tensor_gpr} relies on. They are not viable as the operational kernel for large-$N$ reconstruction; we include them only as accuracy benchmarks to justify the MoG-$3$ choice.

The closed-form expressions and fitted parameters for all five candidate kernels on the JHTDB pressure slice are collected in Table~\ref{tab:kernel_models}.

\begin{table}[h!]
\centering
\label{tab:kernel_models}
\begin{tabular}{l l l}
\toprule
Kernel & $\mathcal{K}(r)$ & Fitted parameters \\
\midrule
Gaussian      & $\exp\!\bigl(-r^{2}/(2\ell^{2})\bigr)$
              & $\ell = 0.234$ \\
Mat\'ern-$3/2$ & $\bigl(1+\sqrt{3}\,r/\ell\bigr)\exp\!\bigl(-\sqrt{3}\,r/\ell\bigr)$
              & $\ell = 0.269$ \\
MoG-$3$       & $\sum_{i=1}^{3}\alpha_i \exp\!\bigl(-r^{2}/(2\ell_i^{2})\bigr),\ \sum_i\alpha_i=1$
              & \begin{tabular}[c]{@{}l@{}}
                  $(\alpha_1,\ell_1) = (0.055, 0.035)$\\
                  $(\alpha_2,\ell_2) = (0.403, 0.242)$\\
                  $(\alpha_3,\ell_3) = (0.542, 0.242)$
                \end{tabular} \\
spectral-clip & $\mathcal{H}^{-1}\!\max(\mathcal{H}\mathcal{K},0)$ (no closed form)
              & --- \\
Bessel-Gauss  & $a_1 e^{-r^{2}/(2\ell_1^{2})} + a_2 J_0(k r)\,e^{-r^{2}/(2\ell_2^{2})},\ a_1+a_2=1$
              & \begin{tabular}[c]{@{}l@{}}
                  $(a_1,\ell_1) = (0.341, 0.161)$\\
                  $(a_2,\ell_2) = (0.659, 0.618)$\\
                  $k = 4.58$
                \end{tabular} \\
\bottomrule
\end{tabular}
\caption{Closed-form expressions for the candidate kernels of Figure~\ref{fig:hyper_optimize}(a) and their fitted parameters on the JHTDB pressure slice. All kernels are normalized to $\mathcal{K}(0)=1$ and scaled by $\sigma_p^{2}$ in use; spectral-clip has no closed form (it is the inverse Hankel transform of $\max(\mathcal{H}\mathcal{K},0)$ on the empirical $\mathcal{K}$).}
\end{table}

\begin{figure}[!h]
    \centering
        \includegraphics[width=0.95\textwidth]{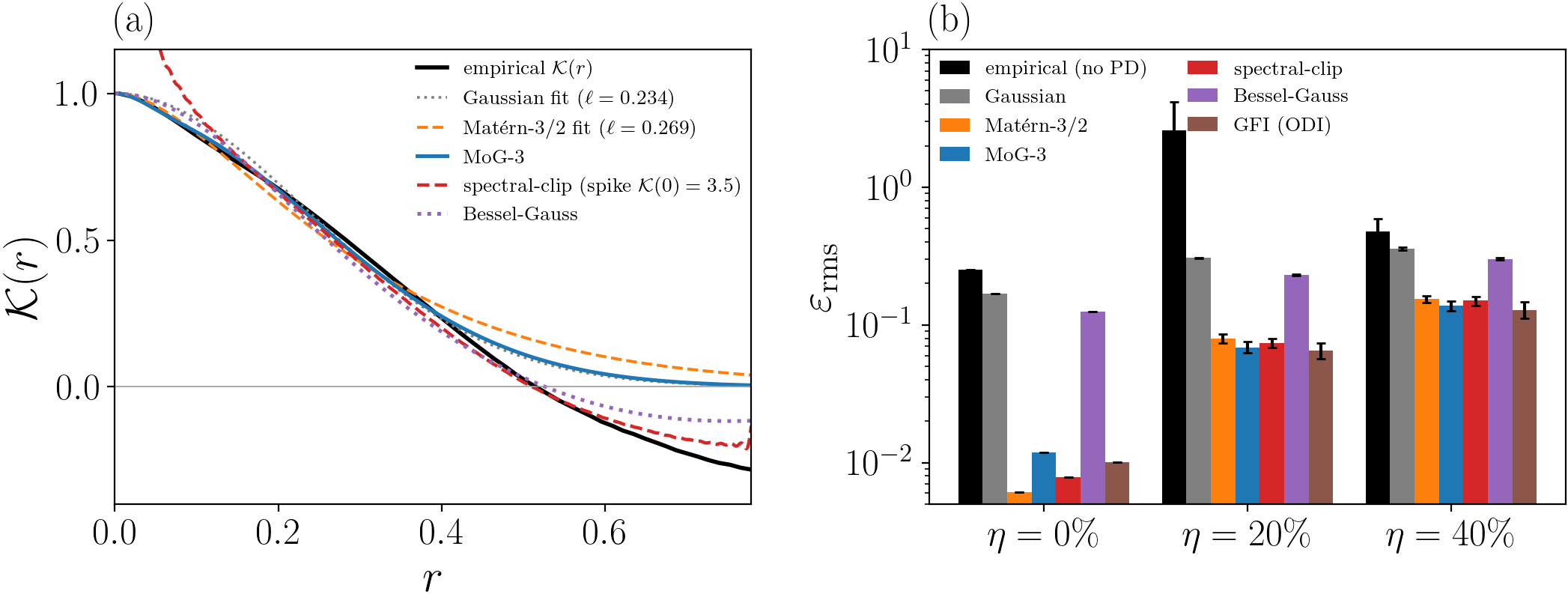}
    \caption{Empirical-kernel design for the 2D GPR reconstruction. (a) Azimuthally averaged correlation function $\mathcal{K}(r)$ of the true pressure field (solid black) with three parametric fits and the non-parametric MoG-$3$ projection adopted in this paper. (b) Relative RMSE $\varepsilon_{\mathrm{rms}}$ of dense-Gram GPR reconstruction using each candidate kernel as the prior covariance, at three gradient-noise levels $\eta\in\{0,\,20\%,\,40\%\}$, on a centered $48\times48$ subgrid of the JHTDB pressure slice. Error bars are over three noise realizations. The same dense-Gram machinery is used for every GPR kernel, including the non-separable spectral-clip and Bessel-Gauss.}
    \label{fig:hyper_optimize}
\end{figure}

To quantify reconstruction accuracy, we use the relative pointwise reconstruction error. The pointwise residual is $\epsilon=p_{rec}-\tilde{p}$, where $\tilde{p}$ denotes the true field and $p_{rec}$ is the reconstructed field. The spatial root-mean-square of the residual on a single realization is denoted as $\epsilon_{\mathrm{rms}}$, and the relative error averaged over $N_s$ noise realizations is denoted $\displaystyle\varepsilon_{\mathrm{rms}} \equiv  \frac{1}{N_s}\sum_{n=1}^{N_s}\!\left(\frac{\epsilon_{\mathrm{rms}}}{\tilde{p}_{\mathrm{rms}}}\right)_{\!n}$,
where $\tilde{p}_{\mathrm{rms}}$ is the spatial standard deviation of the true field. Both $p_{GPR}$ and $\tilde{p}$ are taken with zero spatial mean before the residual is formed, consistent with the zero-mean gauge fixed by the GPR prior in \S\ref{sec:formulation}.

Figure~\ref{fig:hyper_optimize}(a) overlays all five candidate kernels against the empirical $\mathcal{K}(r)$. The Gaussian and Mat\'ern-$3/2$ forms leave a visible residual at moderate $r$ since they have only one length scale; the MoG-$3$ projection matches the near-origin branch almost exactly; spectral-clip concentrates additional mass into a delta-like spike at the origin (the only way to annihilate the oscillatory modes whose Hankel transforms went negative is to redistribute the corresponding spatial energy into a point mass); Bessel-Gauss is the only projection that reproduces the negative lobe, but at the price of a noticeably narrower near-origin peak. Panel (b) reports the relative RMSE $\varepsilon_{\mathrm{rms}}$ of the corresponding GPR reconstruction at $\eta\in\{0,0.2,0.4\}$. We note that the panel-(b) benchmark is run on a centered $48\times48$ subgrid with a dense Gram solve, so that all candidate kernels --- including the non-separable spectral-clip and Bessel-Gauss --- can be tested on identical machinery. The absolute error levels, therefore, differ from the $256\times256$ numbers reported for Figure~\ref{fig:True_GPR_GFI}; the panel is intended as a relative comparison among kernels.

On this benchmark, MoG-$3$ is at least as accurate as the other GPR projections at every noise level tested, and statistically indistinguishable from the $L^{2}$-optimal spectral-clip. The relative ranking of the remaining projections (e.g.\ whether Bessel-Gauss's narrower near-origin peak or its retained negative lobe is the dominant factor) is harder to read off a $48\times48$ subgrid, where finite-domain effects bias the comparison; we therefore avoid drawing strong conclusions about which feature of $\mathcal{K}(r)$ matters most. What is robust on this benchmark is that MoG-$3$ matches the empirical correlation in closed form and uniquely preserves the Kronecker structure required by the fast solver of \S\ref{sec:formulation}; we adopt it as the GPR prior covariance for all subsequent reconstructions in this paper. The accuracy of the resulting scheme against the GFI baseline across noise levels is examined below on the full $256\times256$ slice.

Once the kernel is fixed by the data, the only remaining hyperparameter is the assumed noise level $\sigma_\varepsilon$, which sets the Tikhonov-like regularization of $K_{gg}+\sigma_\varepsilon^{2}I$. We find that $\varepsilon_{\mathrm{rms}}$ is insensitive to $\sigma_\varepsilon$ over more than a decade --- a direct consequence of the spectrum-consistent prior, which makes the Wiener filter near-optimal across noise levels. We therefore fix $\sigma_\varepsilon=\Delta/\sqrt{3}$ (the measured uniform-noise standard deviation) throughout the remainder of \S\ref{sec:problem}.

\subsection{GPR evaluation and uncertainty quantification}
\label{sec:posterior_uq}
\begin{figure}[h!]
    \centering
    \includegraphics[width=0.40\textwidth]{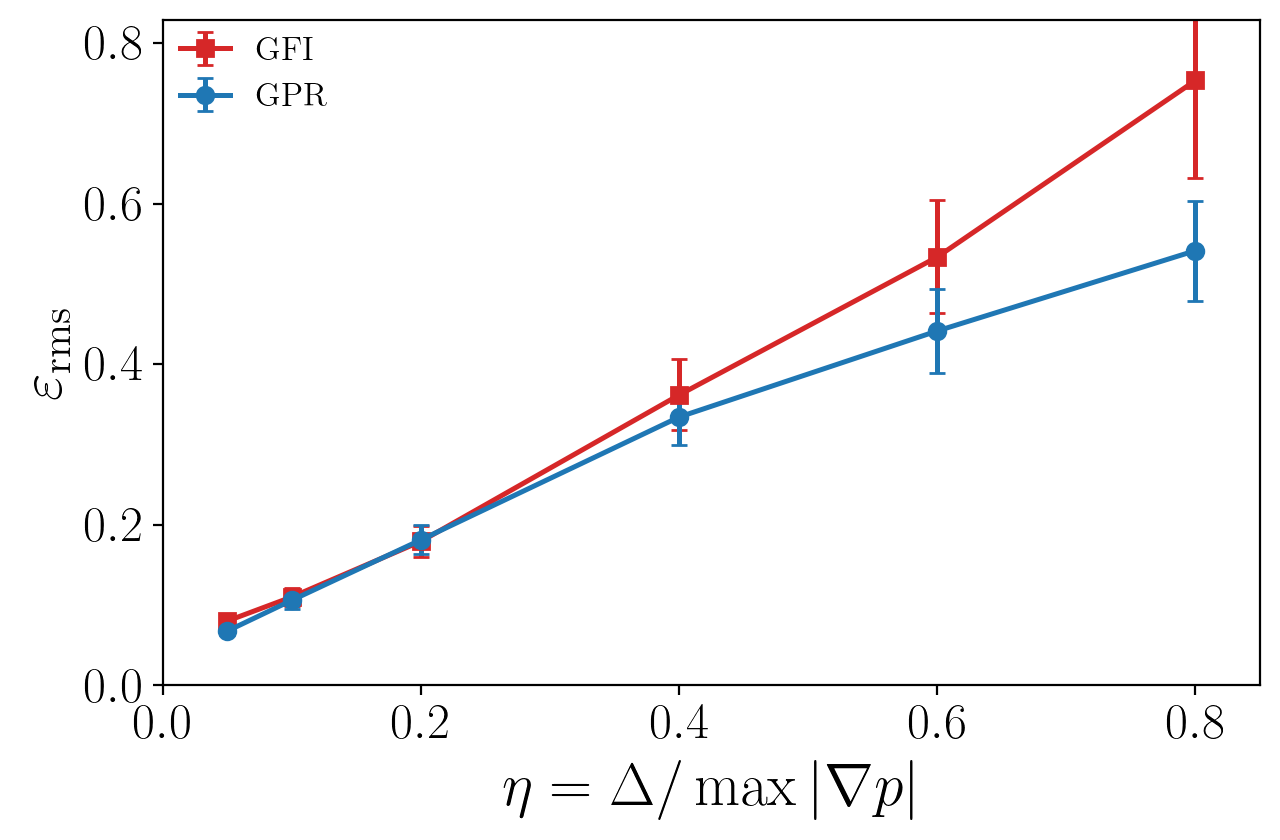}
    \caption{Relative reconstruction RMSE $\varepsilon$ versus gradient-observation noise amplitude $\eta=\Delta/\max|\nabla p|$ on the stride-$4$ subsample of the JHTDB isotropic-turbulence slice ($N=64$, $L=\pi/2$, the same operating point as Figs.~\ref{fig:power_spectrum} and \ref{fig:bias_variance_prediction}(a)). Each marker is the mean of 15 independent uniform-noise realizations with the empirical MoG-$3$ kernel; error bars $\pm\sigma$. Linear axes; $\eta$ runs from $5\%$ to $80\%$, covering the realistic PIV-noise range. Uniform noise on $[-\Delta,\Delta]$ with $\Delta = \eta\,\max|\nabla p|$ is equivalent to Gaussian observation noise of standard deviation $\sigma_\varepsilon = \eta\,\max|\nabla p|/\sqrt{3}$.}
    \label{fig:noise_robustness}
\end{figure}
With both the kernel and $\sigma_\varepsilon$ fixed, the proposed GPR scheme is fully specified and is tested against the GFI baseline. Figure~\ref{fig:noise_robustness} sweeps the gradient-observation noise amplitude $\eta=\Delta/\max|\nabla p|$ from $5\%$ to $80\%$, averaged over fifteen independent uniform-noise realizations at each level. The two methods are statistically tied for $\eta\lesssim 20\%$, reflecting that both GFI and GPR can achieve desired accuracy when provided with full-resolution data with low noise. They separate visibly as $\eta$ grows: at $\eta=60\%$, GPR's relative RMSE is $\approx 25\%$ lower than GFI's ($0.44$ vs $0.53$); at $\eta=80\%$, $\approx 30\%$ lower ($0.54$ vs $0.75$). GFI's curve is essentially linear in $\eta$, whereas GPR's curve bends sub-linearly because the prior-imposed kernel cut limits the noise gain. The trial-to-trial standard deviation grows with $\eta$ for both methods but is consistently smaller for GPR.

In addition, GPR provides a posterior covariance through,
\begin{equation}
    C_{\mathrm{GPR}}(\boldsymbol{X},\boldsymbol{X})
    \;=\; K_{pp} - K_{pg}\bigl(K_{gg}+\sigma_\varepsilon^{2}\mathbf{I}_*\bigr)^{-1}K_{gp}.
\end{equation}
Under the Kronecker tensor-product structure of \S\ref{sec:tensor_gpr}, evaluating $\sigma_{\mathrm{post}}(\boldsymbol{x})=\sqrt{\mathrm{diag}\,C_{\mathrm{GPR}}}$ at a single pixel reduces to one warm-started CG solve of $K_{gg}+\sigma_\varepsilon^{2}I$, and warm-starting the solve from the previous pixel's solution makes the entire $N^{2}$-pixel map cost no more than the original reconstruction. The whole map for the $128^{2}$ benchmark takes about thirty minutes on a single CPU.

\begin{figure}[!h]
    \centering
    \includegraphics[width=0.8\textwidth]{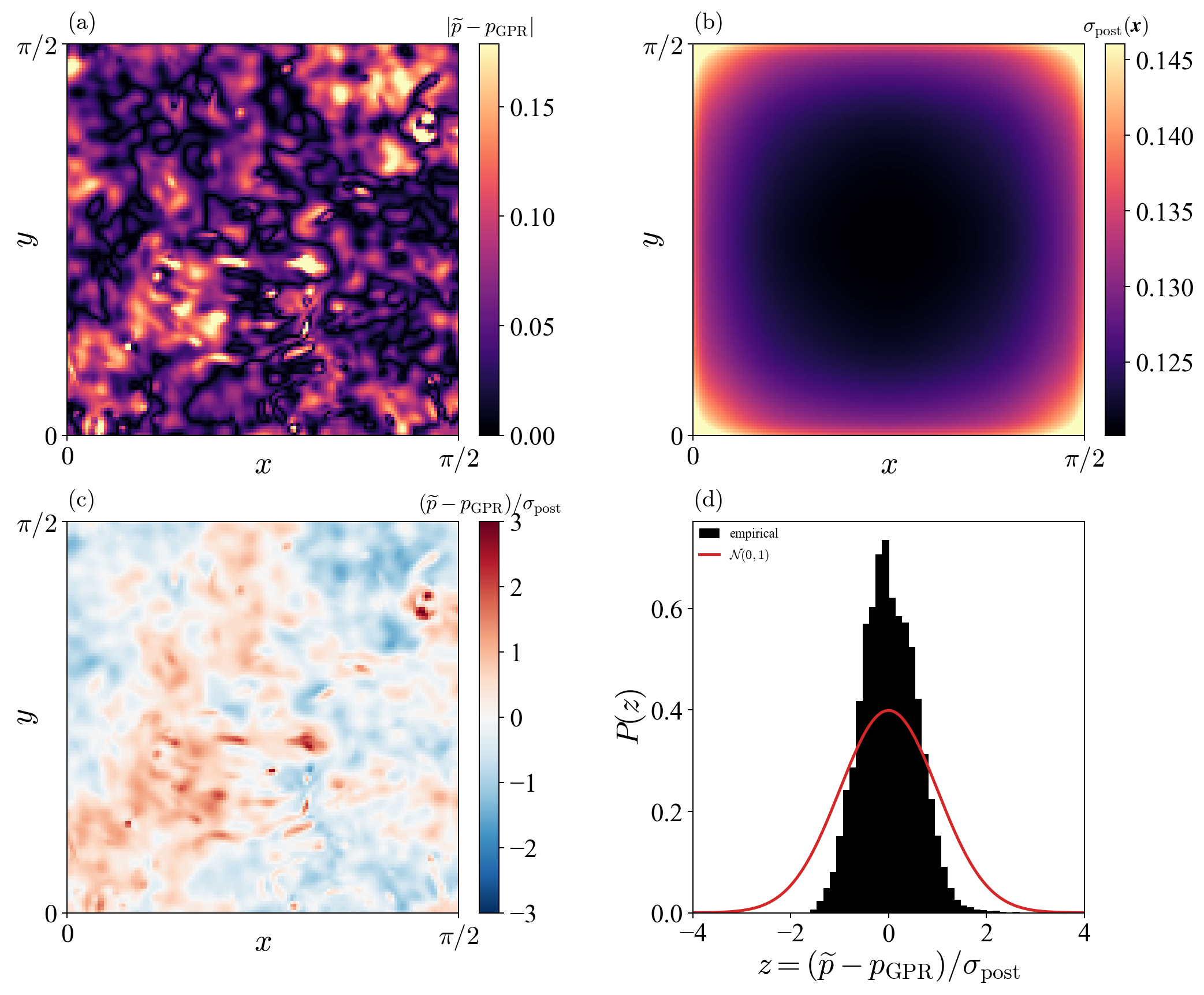}
    \caption{Posterior uncertainty quantification on the 2D JHTDB benchmark at $\eta=40\%$ gradient noise, $N=128$. (a) Magnitude of the mean-subtracted reconstruction error $|\tilde{p}-p_{\mathrm{GPR}}|$, RMS $\varepsilon=0.071$. (b) Pointwise predictive standard deviation $\sigma_{\mathrm{post}}(\boldsymbol{x})$ obtained from the diagonal of the posterior covariance. (c) Standardized residual $z=(\tilde{p}-p_{\mathrm{GPR}})/\sigma_{\mathrm{post}}(\boldsymbol{x})$. (d) Empirical distribution $P(z)$ of $z$ across the grid (black histogram) compared to $\mathcal{N}(0,1)$ (red).}
    \label{fig:posterior_uq}
\end{figure}

Figure~\ref{fig:posterior_uq} plots the error analysis for a sample 2D GPR reconstruction. 
The actual error shown in $(a)$ is highly inhomogeneous and concentrates along the sharp-gradient ridges that the smooth MoG-$3$ prior under-resolves. 
The predicted standard deviation shown in $(b)$ is nearly uniform across the bulk of the domain ($\sigma_{\mathrm{post}}\approx 0.122$) and rises by about $20\%$ to $\sigma_{\mathrm{post}}\approx 0.146$ within roughly one correlation length of the boundary, where the effective observation density seen by the kernel is reduced. 
Comparing the two, the pointwise $\sigma_{\mathrm{post}}$ acts as a conservative envelope for the actual error: the true error sits well within the $\pm\sigma_{\mathrm{post}}$ band almost everywhere, with $|z|<2$ for more than $95\%$ of grid points, as shown in $(c)$ and $(d)$.

\subsection{Impulse-response and spectral analysis}
\label{sec:impulse_response_spectral}

\begin{figure}[tb]
    \centering
    \includegraphics[width=\textwidth]{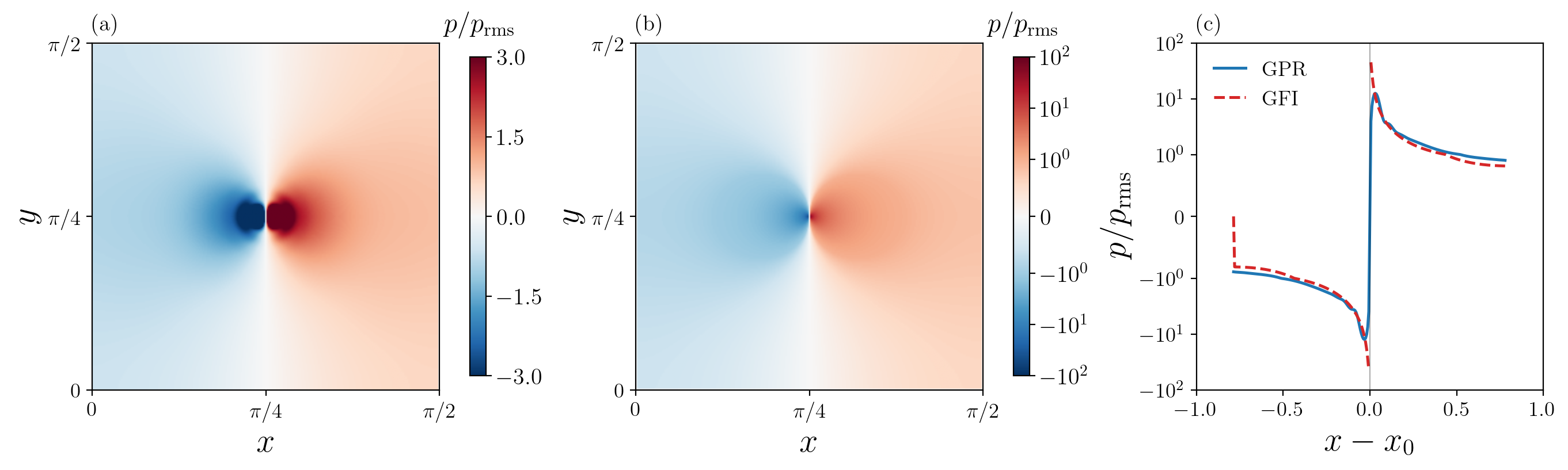}
    \caption{Impulse response of the GPR and GFI inverse operators to a unit-amplitude single-pixel $\partial p/\partial x$ perturbation at the center of the $256\times 256$ JHTDB slice ($\partial p/\partial y$ held to zero). The GPR operator $\mathbf{R}=K_{pg}(K_{gg}+\sigma_\varepsilon^{2}\mathbf{I})^{-1}$ is solved by conjugate gradients on the full Gram matrix; the noise jitter is set at the $\eta=40\%$ Gaussian-equivalent operating point of \S\ref{sec:posterior_uq} ($\sigma_\varepsilon=0.40\,\max|\nabla\tilde p|/\sqrt{3}\approx 8.94$ in the JHTDB units), so the figure depicts the operator at the same regime as the noise-robustness study. GFI is deterministic and uses no jitter. Each field is normalized by its own root-mean-square value $p_{\mathrm{rms}}=\langle p^2\rangle^{1/2}$. (a) GPR response with the empirical MoG-$3$ kernel on a linear color scale. (b) GFI response on a symmetric-log color scale. (c) Mid-domain cross-sections $p(x, \pi/4)$ of the GPR (blue) and GFI (red dashed) responses on a symlog vertical axis with the same normalization.}
    \label{fig:impulse_response}
\end{figure}

To probe the spatial footprint of the GPR and GFI operators, we perform the domain-of-influence diagnostic \citep{wang_wang_zaki_2022,wang2019spatial} and evaluate each reconstruction on a single-point impulse in the observed $x$-gradient at the center of the domain, with the $y$-gradient held at zero. The resulting pressure fields in Figure~\ref{fig:impulse_response}(a,b) are the impulse response of the two operators.

The two responses share the same antisymmetric dipole structure across $\boldsymbol{x}_0$, in agreement with the discussion in \S\ref{sec:GPR_GFI_connection}. The impulse responses far away from the perturbation and away from the boundaries are nearly identical. They depart mainly near the perturbation. GFI inherits the $\log r$ singularity of the 2D Laplacian's free-space Green's function, which is regularized by GPR. Panel (c) makes the regularization explicit. The figure visually realizes that GPR is a kernel-regularized GFI, producing the noise-robustness gain of Figure~\ref{fig:noise_robustness}.

The same denoising-vs-fidelity contrast is most cleanly diagnosed in the wavenumber domain. With $\widehat{p}(\boldsymbol{k})$ the discrete Fourier transform of an $N_x\!\times\!N_y$ pressure field, the radial pressure spectrum is the standard annular-shell average
\begin{equation}
    E_p(|\boldsymbol{k}|_b)
      = \frac{1}{\Delta k}
        \sum_{|\boldsymbol{k}|_b^\ominus \le |\boldsymbol{k}| < |\boldsymbol{k}|_b^\oplus}
          \bigl|\hat{p}(\boldsymbol{k})\bigr|^{2},
    \label{eq:radial_spectrum}
\end{equation}
binned to the Nyquist magnitude with $\Delta k$-wide shells and Parseval-normalized so that $\sum_b E_p\,\Delta k = \operatorname{Var}(p)$; the zero-frequency bin is discarded.

\begin{figure}[tb]
    \centering
    \includegraphics[width=\textwidth]{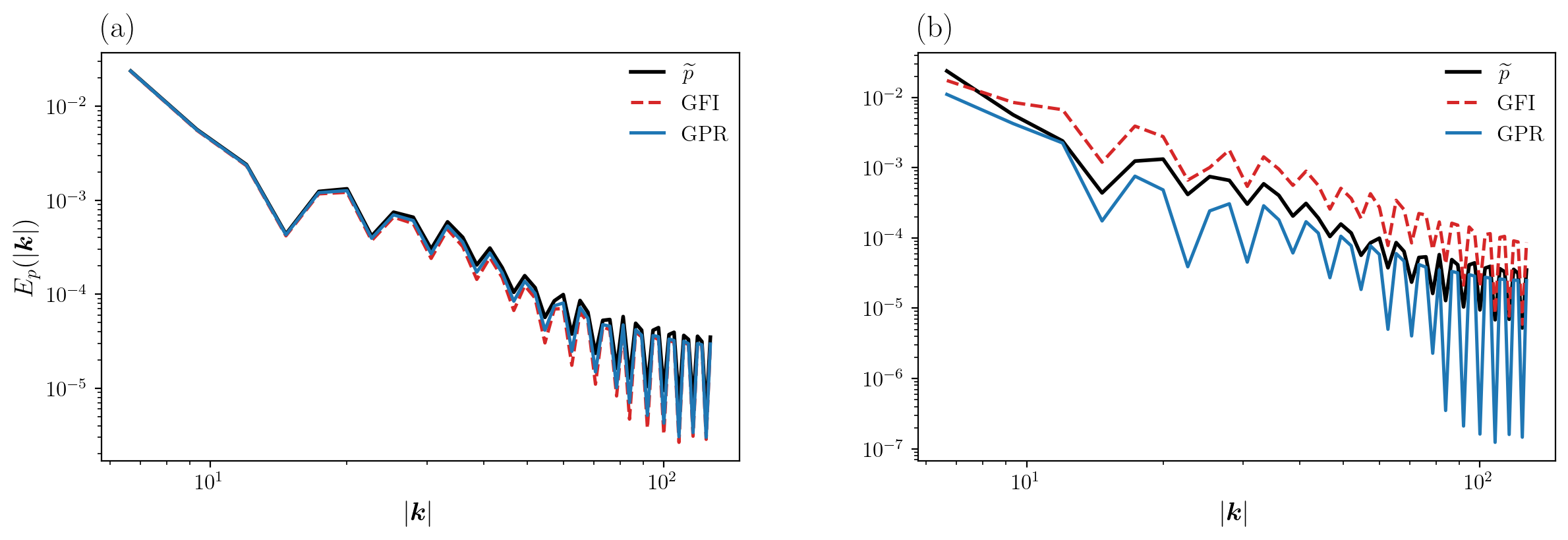}
    \caption{Radial energy spectrum $E_p(|\boldsymbol{k}|)$ of the true pressure field (black), the GFI reconstruction (red dashed), and the GPR reconstruction (blue), computed by the annular-shell average of Eq.~\eqref{eq:radial_spectrum} on the stride-$4$ subsample of the JHTDB slice ($N=64$, $L=\pi/2$). (a) Exact gradient observations ($\eta=0$). (b) error-embedded observations at $\eta=80\%$. The GFI noise shelf is clearly visible above the truth at intermediate-to-high $|\boldsymbol{k}|$.}
    \label{fig:power_spectrum}
\end{figure}
Figure~\ref{fig:power_spectrum}(a) shows that with exact gradient data, both reconstructions recover the truth across the full wavenumber range.
Figure~\ref{fig:power_spectrum}(b) plots the same comparison at $\eta=80\%$. At intermediate-to-high $|\boldsymbol{k}|$, the GFI spectrum lifts above the truth, consistent with the GFI operator's $1/|\boldsymbol{k}|$ Fourier symbol propagating gradient noise unfiltered into pressure, while the GPR spectrum sits modestly below the truth, over-attenuating the error, which reflects the low-pass character of the empirical kernel. The integrated error of GFI dominates that of GPR by a factor of $1.4$. The trade-off between these two failure modes is precisely the per-mode Wiener filter that the spectrum-consistent prior implements: by Bochner's theorem the empirical-correlation MoG-$3$ fit is equivalent to a fit of the empirical pressure spectrum $S(\boldsymbol{k})$, and on an idealized periodic domain the GPR posterior reduces analytically to $\widehat{p}[\boldsymbol{k}] = S/(\sigma_\varepsilon^{2} + S\|\boldsymbol{k}\|^{2})\,(-i\boldsymbol{k}\cdot\widehat{\boldsymbol{O}})$, which saturates the linear-estimator Wiener floor by construction.

The denoising behavior visible in Figures~\ref{fig:impulse_response} and \ref{fig:power_spectrum} originates in the singular-value structure of the GPR reconstruction operator $\mathbf{R} \equiv K_{pg}(K_{gg}+\sigma_\varepsilon^{2}\mathbf{I}_*)^{-1}$, which maps the stacked gradient observations to the reconstructed pressure. For i.i.d.\ gradient noise of variance $\sigma_\varepsilon^{2}$, the per-pixel reconstruction noise variance averages to $\sigma_\varepsilon^{2}\,(N_xN_y)^{-1}\!\sum_m \lambda_m^{2}$, with $\{\lambda_m\}$ the singular values of $\mathbf{R}$: rapid decay of $\{\lambda_m\}$ implies strong noise suppression at the cost of attenuating genuine high-wavenumber content. The rest of this section examines $\{\lambda_m\}$ directly.

\begin{figure}[tb]
    \centering
    \includegraphics[width=10 cm]{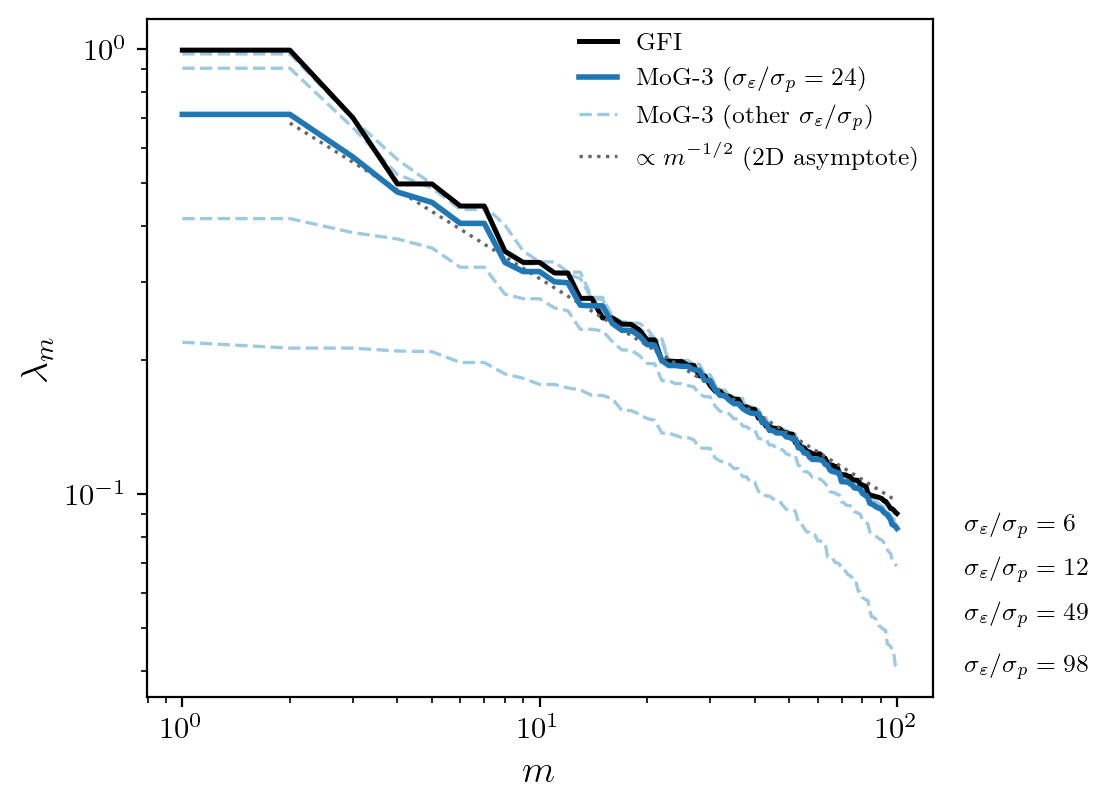}
    \caption{Singular values $\lambda_m$ of the GFI (black, solid) and MoG-$3$ GPR (blue, solid) operators on a $256\times 256$ cell grid spanning $L=\pi$, at the benchmark operating point $\sigma_\varepsilon/\sigma_p\approx 24$ (equivalent to the $\eta=40\%$ gradient-noise level). Light-blue dashed curves show MoG-$3$ at additional $\sigma_\varepsilon/\sigma_p\in\{98, 49, 12, 6\}$ to illustrate the convergence to the GFI spectrum as $\sigma_\varepsilon\to 0$.}
    \label{fig:svd_values}
\end{figure}

For the SVD analysis below we use a single domain size $L=\pi$ (about $13$ correlation lengths per side, taking the kernel correlation length to be $\ell_c\approx 0.25$, the half-decay point read off the empirical $\mathcal{K}(r)$ in Figure~\ref{fig:hyper_optimize}(a)), larger than the main 2D benchmark of Table~\ref{tab:2d_setup} ($L=\pi/2$). 
The domain is large enough to contain an asymptotic GFI decay law in Figure~\ref{fig:svd_values}. We have verified that the qualitative spectral picture is preserved at other domain sizes (e.g.\ $L=2\pi$).

Figure~\ref{fig:svd_values} shows the operating-point spectra on the $256\times 256$ cell grid at $L=\pi$. With the gauge projected out, the GFI spectrum peaks at $\lambda_1\approx 1.0$ (the twofold-degenerate dipolar mode) and then decays algebraically as $\lambda_m\propto m^{-1/2}$ over the resolved range. The $m^{-1/2}$ rate is dictated by the 2D structure of the GFI kernel: $G(\boldsymbol{r}) = -\log|\boldsymbol{r}|/(2\pi)$ is the inverse-Laplacian Green's function, so the gradient-to-pressure operator has Fourier symbol $|\nabla G(\boldsymbol{k})|=1/|\boldsymbol{k}|$. On a 2D box the number of distinct $\boldsymbol{k}$-modes inside a disc scales with the square of the radius, so the $m$-th sorted singular value sits at radial wavenumber $|\boldsymbol{k}|\propto\sqrt{m}$ and inherits the scaling $\lambda_m\propto 1/|\boldsymbol{k}|\propto m^{-1/2}$. The dotted reference line in Figure~\ref{fig:svd_values} is this 2D asymptote.

MoG-$3$'s leading value is smaller than GFI's, reflecting the kernel prior's additional shrinkage across the full operator. The MoG-$3$ spectrum for $\sigma_{\varepsilon}/\sigma_p = 98$ is nearly flat up to $m\sim 10$: in this band the kernel still resolves the corresponding wavenumbers, the Wiener regularizer $1/(K_{gg}+\sigma_\varepsilon^{2})$ is dominated by $K_{gg}\!\gg\!\sigma_\varepsilon^{2}$, and the operator essentially reduces to the inverse-Laplacian sub-symbol of GFI, scaled down by the kernel-implied prior variance. Past the kernel-resolved bandwidth, the spectrum of the GPR operator falls off with an exponential decay. This is the spectral signature of the difference in noise-suppression behavior between the two operators: GFI inherits the unbounded high-wavenumber response of the bare 2D Green's function, whereas MoG-$3$'s prior provides a built-in low-pass cut at $|\boldsymbol{k}|\!\sim\!1/\ell_c$ that bounds the high-frequency contribution to the noise gain regardless of $N$. This is the operator-level origin of the denoising advantage.

\begin{figure}[tb]
    \centering
    \includegraphics[width=\textwidth]{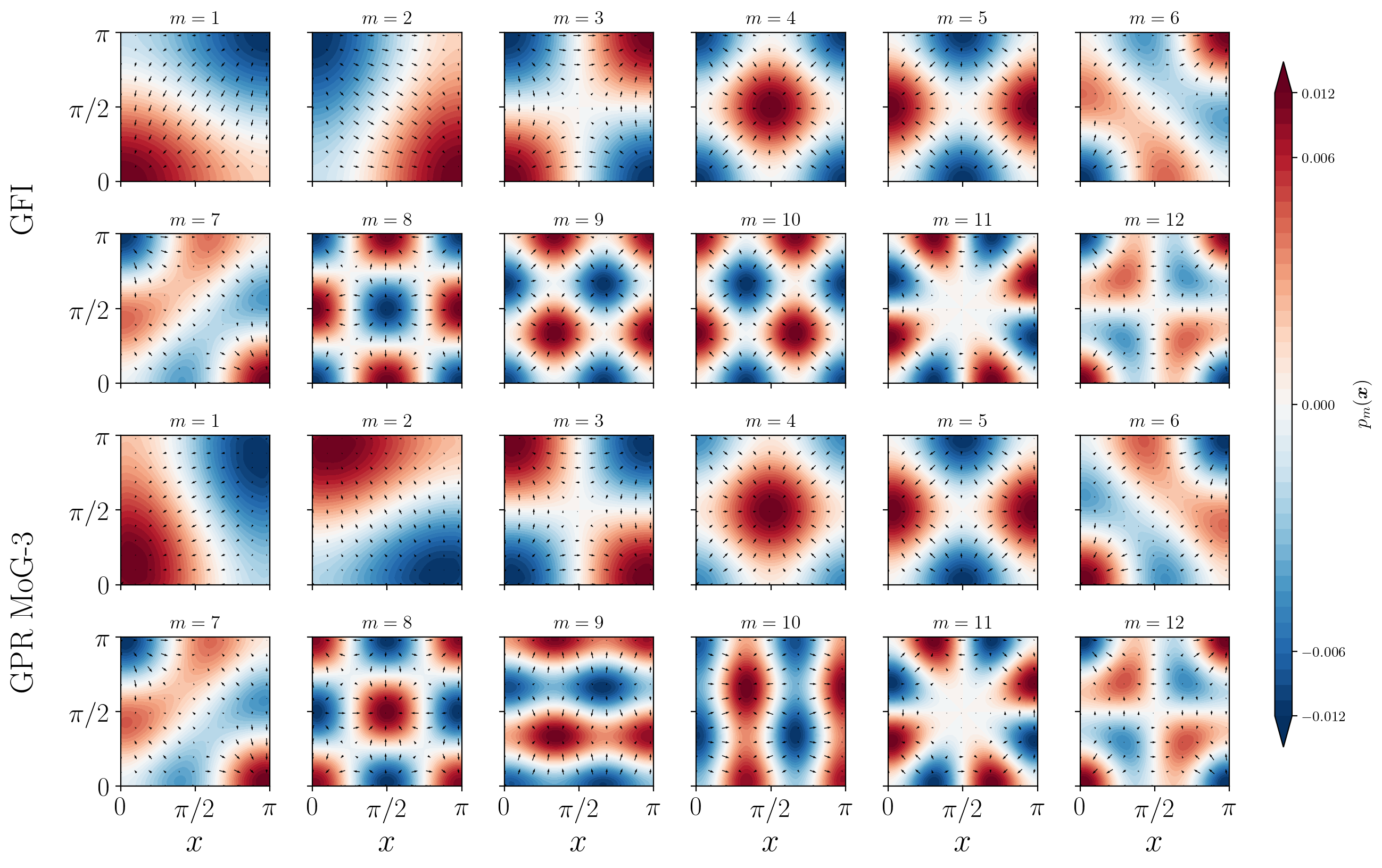}
    \caption{Leading twelve singular modes of the GFI (rows 1--2) and MoG-$3$ GPR (rows 3--4) operators on a $256\times 256$ cell grid spanning $L=\pi$. The noise level is $\sigma_\varepsilon/\sigma_p\approx 24$ (matching the $\eta=40\%$ benchmark). Each row shows six consecutive modes; each panel combines the pressure mode $u_k(\boldsymbol{x})$ as a filled contour with the associated gradient mode $v_k(\boldsymbol{x})$ overlaid as a quiver (subsampled to $\sim 10$ arrows per side), both per-panel normalized.}
    \label{fig:svd_modes}
\end{figure}

As a reference, on a periodic square of side $L$ both operators would be translation-invariant and would diagonalize simultaneously in the Fourier basis $\{\exp(\mathrm{i}\boldsymbol{k}\cdot\boldsymbol{x})\}$, with
\begin{equation}
    \lambda_{\boldsymbol{k}}^{\mathrm{GFI}} = \frac{1}{|\boldsymbol{k}|},
    \qquad
    \lambda_{\boldsymbol{k}}^{\mathrm{GPR}}
        = \frac{|\boldsymbol{k}|\,\sigma_p^{2}\widehat{\mathcal{K}}(\boldsymbol{k})}
               {|\boldsymbol{k}|^{2}\,\sigma_p^{2}\widehat{\mathcal{K}}(\boldsymbol{k})
                + \sigma_\varepsilon^{2}}.
                \label{eq:svd_analytic}
\end{equation}
Equation~\eqref{eq:svd_analytic} qualitatively explains the empirical observations of Figure~\ref{fig:svd_values}: at low $|\boldsymbol{k}|$, where $|\boldsymbol{k}|^{2}\sigma_p^{2}\widehat{\mathcal{K}}\!\gg\!\sigma_\varepsilon^{2}$, the two symbols coincide as $\lambda_{\boldsymbol{k}}^{\mathrm{GPR}}\!\to\! 1/|\boldsymbol{k}| = \lambda_{\boldsymbol{k}}^{\mathrm{GFI}}$, accounting for the $m^{-1/2}$ asymptote across the kernel-resolved band; at high $|\boldsymbol{k}|$, $\widehat{\mathcal{K}}$ falls off as a Gaussian and the GPR symbol is suppressed multiplicatively by $\sigma_p^{2}\widehat{\mathcal{K}}/\sigma_\varepsilon^{2}\!\to\! 0$, producing the supra-algebraic tail past the kernel-resolved bandwidth. 

Figure~\ref{fig:svd_modes} visualizes the leading twelve pressure and gradient modes of both reconstruction operators at the same $L=\pi$ used for the spectrum. Both operators show a near-identical progression of mode shapes. This is not coincidental: both operators admit closed-form singular value decompositions on an idealized periodic square that the actual finite-domain SVD inherits qualitatively away from the boundary. The numerical operators considered here are not periodic, so the formulas below describe the bulk symbol, not exact eigenvalues.

The structural connection between this spectral picture and the Green's-function-integration baseline introduced in \S\ref{sec:GPR_GFI_connection} is then complete: the GPR operator's symbol \eqref{eq:svd_analytic} reduces exactly to GFI's $1/|\boldsymbol{k}|$ in the noiseless, kernel-resolved limit, and departs from it only where the empirical kernel power $\sigma_p^{2}\widehat{\mathcal{K}}(\boldsymbol{k})$ has rolled off below $\sigma_\varepsilon^{2}/|\boldsymbol{k}|^{2}$.

% -----------------------------

% -----------------------------

\section{Three-dimensional pressure reconstruction}
\label{sec:3Dresults}
GPR has been applied to three-dimensional reconstruction problems in several contexts.
In plasma physics, \citet{Howell_Hanson_2020} implemented a non-parametric GPR model within the V3FIT code for three-dimensional equilibrium reconstruction, combining GPR with parametric techniques to infer equilibrium profiles from experimental data.
\citet{yang2024fast} developed a reduced-order model coupling Proper Orthogonal Decomposition (POD) with an improved Automatic Kernel Construction GPR (AKC-GPR) for predicting 3D flow fields around hypersonic vehicles.
While these studies demonstrate the versatility of GPR in 3D settings, none address the integration-constant problem that arises when applying GPR plane-by-plane for pressure reconstruction from gradient data.

The two three-dimensional strategies introduced in \S\ref{sec:3d_recon} --- plane-wise reconstruction with least-squares integration constants (\S\ref{sec:3d_recon:planewise}) and the tensor-product Kronecker formulation (\S\ref{sec:tensor_gpr}) --- are benchmarked against one another in \S\ref{sec:3d_tensor_comparison} on a JHTDB subvolume, and the impulse response of the 3D operator is examined in \S\ref{sec:3d_impulse}.

\subsection{Comparison with tensor-product three-dimensional GPR}
\label{sec:3d_tensor_comparison}

The plane-wise-plus-least-squares reconstruction reduces a three-dimensional inference to $3N$ two-dimensional solves, at the cost of introducing $3N$ free integration constants and breaking the cross-plane noise correlations implied by the kernel. The tensor-product GPR of \S\ref{sec:tensor_gpr} retains full cross-plane coupling while matching the plane-wise asymptotic cost. In this section, we compare the two formulations on a JHTDB volume at the paper's benchmark noise level.

The test domain is a $64^3$ subregion of the \texttt{isotropic1024coarse} dataset, evaluated at two physical extents to confirm the cross-over noise level is a kernel-resolution property rather than a domain-size artifact: panel~(a) at $L=\pi/8$ and panel~(b) at $L=\pi/2$. Pressure gradients are corrupted with uniform noise of amplitude $\eta$ ranging from $2.5\%$ to $80\%$ of $\max|\nabla p|$; the equivalent Gaussian standard deviation is $\sigma_\varepsilon = \eta\,\max|\nabla p|/\sqrt{3}$. Both methods use the same empirical MoG-$3$ kernel re-fitted to each cube; only the inference procedure differs. The tensor-product system is solved with algorithm \ref{alg:tensor_gpr}.

\begin{figure}[h!]
    \centering
    \includegraphics[width=\textwidth]{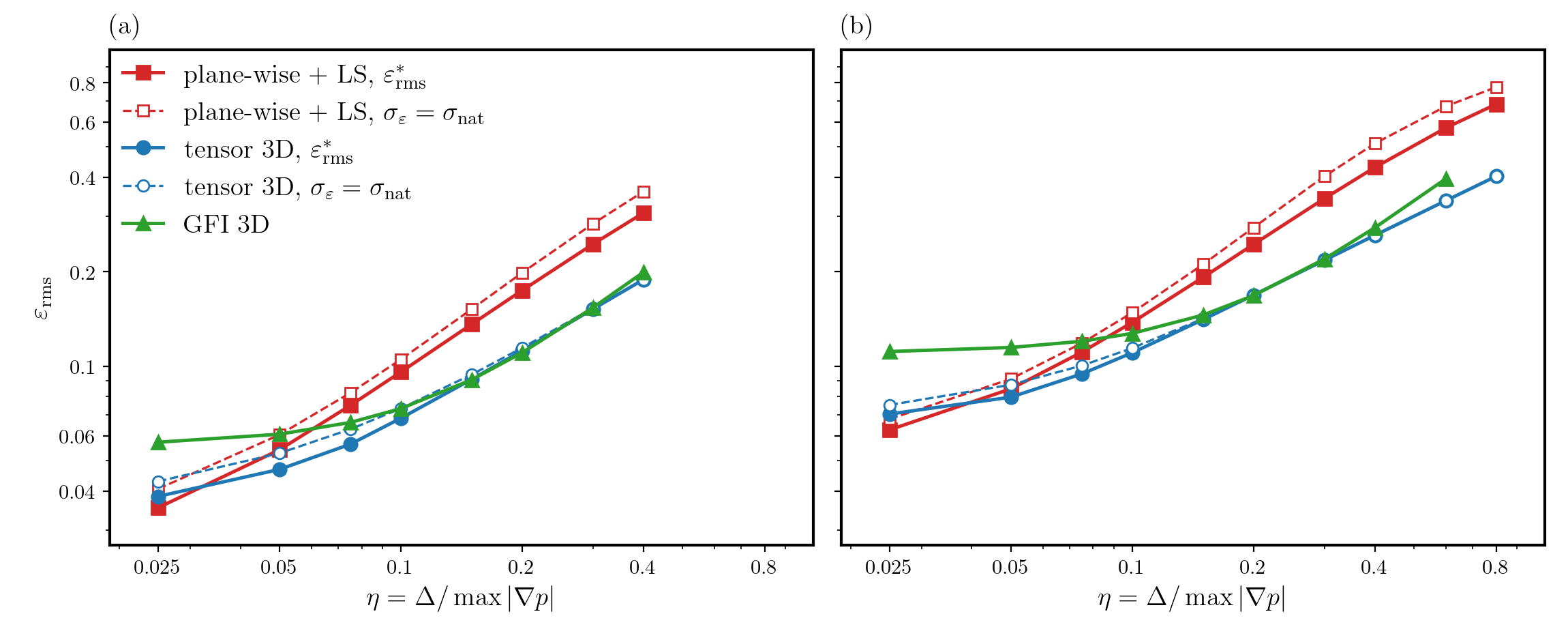}
    \caption{The relative RMSE $\varepsilon_{\mathrm{rms}}$ vs gradient noise $\eta$ on $64^3$ \texttt{isotropic1024coarse} subcubes at two domain extents: (a) $L=\pi/8, N = 128$; (b) $L=\pi/2, N = 256$. Curves are plotted for plane-wise + LS (red squares), tensor-product 3D GPR (blue circles)---both with the same empirical MoG-$3$ kernel re-fitted to each cube---and dense 3D GFI (green triangles) as a hyperparameter-free reference. For each GPR method, the filled solid curve is the best-case $\varepsilon^\star=\min_{\sigma_\varepsilon}\varepsilon$ obtained by scanning $\sigma_\varepsilon$ over multiples of the natural value $\sigma_\varepsilon^{\mathrm{nat}}=\eta\max|\nabla p|/\sqrt 3$, and the open dashed curve is the principled baseline $\varepsilon$ at $\sigma_\varepsilon=\sigma_\varepsilon^{\mathrm{nat}}$.}
    \label{fig:3d_sigma_sweep}
\end{figure}
Figure~\ref{fig:3d_sigma_sweep} reports the RMSE $\varepsilon_{\mathrm{rms}}$ across the noise sweep for three reconstructions: plane-wise + LS, tensor-product 3D GPR (both with the same empirical MoG-$3$ kernel), and dense 3D GFI as a hyperparameter-free reference. For each GPR method, we plot two curves. The filled solid curve is the best-case $\varepsilon^\star_{\mathrm{rms}} = \min_{\sigma_\varepsilon} \varepsilon_{\mathrm{rms}}$, obtained by scanning $\sigma_\varepsilon$ over multiples of the natural value $\sigma_\varepsilon^{\mathrm{nat}}=\eta\max|\nabla p|/\sqrt 3$ implied by the uniform-noise model. The open dashed curve is the principled baseline $\varepsilon_{\mathrm{rms}}$ at $\sigma_\varepsilon=\sigma_\varepsilon^{\mathrm{nat}}$, the value a practitioner would pick from the data alone. On the smaller cube (panel~a) the tensor 3D method is the best across most of the sweep: at low noise ($\eta\le 0.10$) it leads GFI by $20\!-\!30\%$, and is comparable with GFI in the intermediate noise range $\eta\in[0.15,0.20]$, where the kernel-imposed bias and the GFI noise propagation are roughly equal. For $\eta\gtrsim 0.30$, GPR further outperforms GFI by virtue of its low-pass filter properties. Plane-wise + LS is competitive only at the lowest noise. On the larger cube (panel~b), GPR 3D demonstrates a similar advantage. This larger-cube behavior parallels the 2D high-noise behavior of Fig.~\ref{fig:noise_robustness} on the same physical extent. Plane-wise + LS is the worst of the three across most of the sweep.

\begin{figure}[h!]
    \centering
    \includegraphics[width=14 cm]{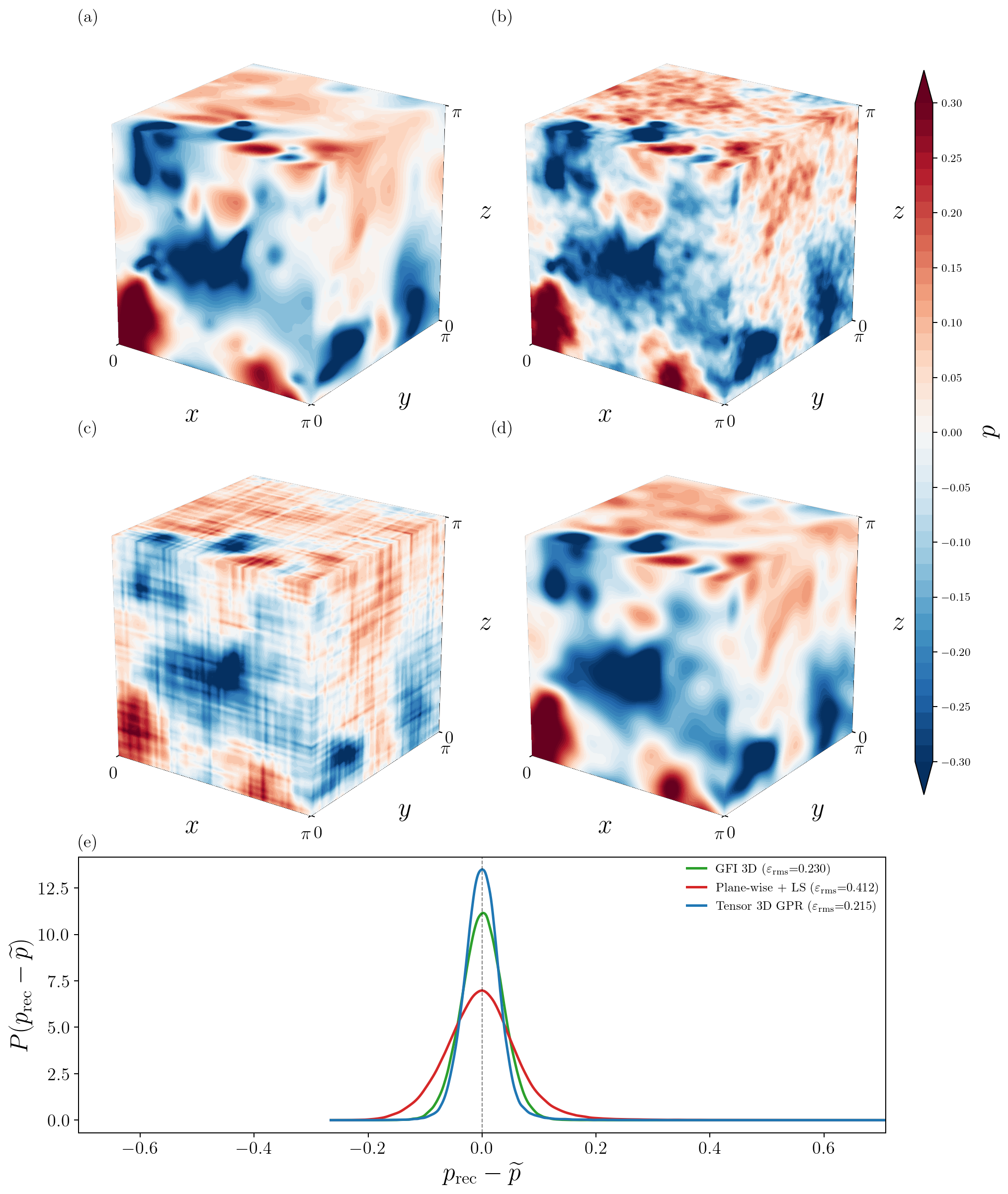}
    \caption{Cube-surface comparison at $40\%$ gradient noise on a JHTDB $64^3$ \texttt{isotropic1024coarse} subcube using the empirical MoG-$3$ kernel for GPR. Each cube shows the pressure on the three visible outer faces. (a) true pressure, (b) 3D GFI reconstruction, (c) plane-wise + LS reconstruction, (d) tensor-product 3D GPR reconstruction, (e) full-volume error probability density functions for the three reconstructions. $\varepsilon_{\mathrm{rms}}$ denotes the relative RMSE.}
    \label{fig:tensor_vs_planewise}
\end{figure}
The optimal ratio $\sigma_\varepsilon^\star/\sigma_\varepsilon^{\mathrm{nat}}$ that achieves the $\varepsilon_{\mathrm{rms}}^\star$ curves above sits at $\approx 0.1$ for both methods at low noise ($\eta\le 0.1$), rising to $\approx 0.5$ in the intermediate range, and approaching $\approx 1$ for the tensor method at moderate-to-high $\eta$ on the larger cube. This is the same flat-basin behavior documented in 2D (Figure~\ref{fig:hyper_optimize}b): $\sigma_\varepsilon$ plays the dual role of noise model and Tikhonov regulariser, and when the MoG-$3$ prior over-smooths the high-frequency tail of the field's spectrum, under-claiming the noise lets the prior do less filtering and yields a modest (10--25\%) MSE improvement.

Figure~\ref{fig:tensor_vs_planewise} visualizes the three reconstructions at $\eta=40\%$ with the natural assumed noise $\sigma_\varepsilon=\sigma_\varepsilon^{\mathrm{nat}}$. Panels (a)--(d) render the pressure on the three visible outer faces of the $64^3$ cube. The plane-wise reconstruction shows a pronounced plane-axis-aligned striping pattern across every face: on an $xy$ face, the contribution from the $xz$- and $yz$-plane solves brings $j$- and $i$-dependent noise from each of the $N$ independent 2D plane solves (each with its own noise realization), which the single additive offset per plane cannot remove. The dense 3D GFI reconstruction has no such striping artifact but inherits the same broadband-noise floor familiar from the 2D analysis, and so sits between the plane-wise baseline and the tensor-product reconstruction. The tensor-product GPR reproduces the fine-scale structure of the true field much more faithfully than either, because it averages noise against the full three-dimensional neighborhood in a single joint inference and applies the empirical MoG-$3$ prior at every wavenumber. Panel (e) shows the full-volume error probability density functions: the tensor method's distribution is the narrowest and least heavy-tailed, GFI is intermediate, and plane-wise + LS is the broadest, consistent with the relative RMSE values.

\subsection{Three-dimensional impulse-response analysis}
\label{sec:3d_impulse}

\begin{figure}[h!]
    \centering
    \includegraphics[width=0.7\textwidth]{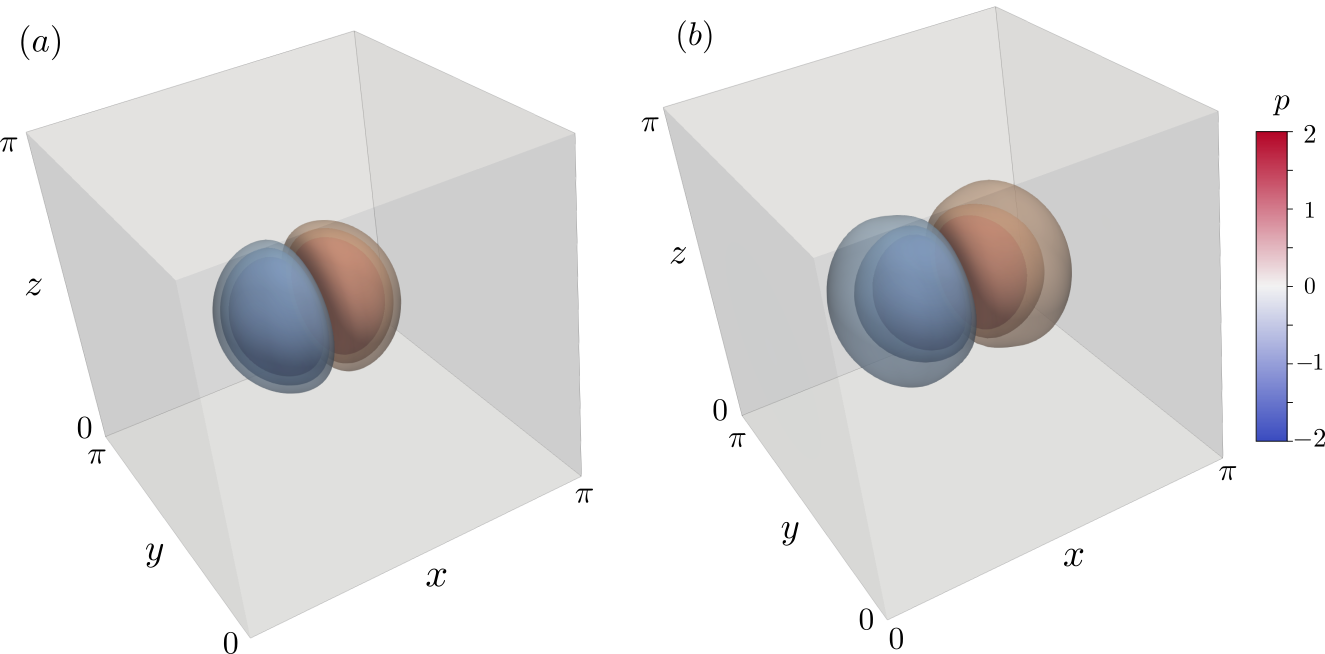}
    \caption{Three-dimensional iso-surface rendering of the (a) GFI and (b) GPR (MoG-$3$)  impulse responses to a single-pixel $\partial_x p$ source at the box center, on $\Omega=[0,\pi]^{3}$, $N=64$. The parameter $\sigma_\varepsilon$ is set to $\sigma_\varepsilon=0.40\,\max|\nabla\tilde p|/\sqrt{3}\approx 19.6$. Each field is normalized by its own RMS $p_{\mathrm{rms}}$ and rendered with the same iso levels $p/p_{\mathrm{rms}}=\pm 0.5,\pm 1,\pm 2$ and the same linear color scale. Both panels exhibit the same antisymmetric dipole structure across $\boldsymbol{x}_{0}$.}
    \label{fig:impulse_response_3d_iso}
\end{figure}

We repeat the 2D impulse-response diagnostic of \S\ref{sec:impulse_response_spectral} in three-dimensional space. A single-pixel $\partial_x p$ impulse is placed at the box center $\boldsymbol{x}_{0}=L/2\cdot(1,1,1)$ on $\Omega=[0,\pi]^{3}$, $N=64$, and reconstructed by the canonical GFI solver and by the full tensor-product MoG-$3$ GPR operator $\mathbf{R}=K_{pg}(K_{gg}+\sigma_\varepsilon^{2}\mathbf{I})^{-1}$. Figure~\ref{fig:impulse_response_3d_iso} renders both responses as iso-surfaces in the same cube using the normalized levels $p/p_{\mathrm{rms}}=\pm 0.5,\pm 1,\pm 2$. The two responses share the same antisymmetric dipole structure, confirming in three dimensions the operator-symbol picture of \S\ref{sec:GPR_GFI_connection}: GPR agrees with GFI on the impulse-response shape, with necessary regularization for denoising effects, manifested in extended structures in panel (b). 
This visualization further confirms the discussion in \S\ref{sec:GPR_GFI_connection}.

\subsection{Predictive scaling from the GPR operator symbol}
\label{sec:bias_variance_prediction}

The trade-off captured in Figure~\ref{fig:3d_sigma_sweep} can be derived directly from the operator symbol of \S\ref{sec:impulse_response_spectral}, without running any reconstruction. We adopt the Gaussian noise model $\boldsymbol{\eta}\sim\mathcal{N}(0,\sigma_\varepsilon^{2}I)$ throughout this subsection (see Table~\ref{tab:2d_setup} for the conversion to the uniform noise used in Figs.~\ref{fig:noise_robustness}--\ref{fig:3d_sigma_sweep}). On a periodic cube of side $L$ with separable kernel and discrete wavenumbers $\boldsymbol{k}\in (2\pi/L)\,\mathbb{Z}^{d}$, the GPR posterior is diagonal in the Fourier basis: substituting the forward model $\widehat{\boldsymbol{g}}(\boldsymbol{k})=\mathrm{i}\boldsymbol{k}\,\widehat{\tilde p}(\boldsymbol{k})+\widehat{\boldsymbol{\eta}}(\boldsymbol{k})$ into Eq.~\eqref{eq:svd_analytic} gives
\begin{equation}
    \varepsilon_{\mathrm{rms}}^{2}\,\sigma_p^{2}
      \;=\;\underbrace{\sum_{\boldsymbol{k}}B(\boldsymbol{k})^{2}\,|\widehat{\tilde p}(\boldsymbol{k})|^{2}}_{\text{signal-attenuation bias}}
      \;+\;\underbrace{\frac{\sigma_\varepsilon^{2}}{N^{d}}\sum_{\boldsymbol{k}}W(\boldsymbol{k})^{2}}_{\text{noise transfer (variance)}},
    \label{eq:bias_var}
\end{equation}
with the Wiener factors
\begin{equation}
    B(\boldsymbol{k})=\frac{\sigma_\varepsilon^{2}}{|\boldsymbol{k}|^{2}\sigma_p^{2}\widehat{\mathcal{K}}(\boldsymbol{k})+\sigma_\varepsilon^{2}},\qquad
    W(\boldsymbol{k})=\frac{|\boldsymbol{k}|\,\sigma_p^{2}\widehat{\mathcal{K}}(\boldsymbol{k})}{|\boldsymbol{k}|^{2}\sigma_p^{2}\widehat{\mathcal{K}}(\boldsymbol{k})+\sigma_\varepsilon^{2}},
    \label{eq:wiener_factors}
\end{equation}
$\widehat{\mathcal{K}}$ the Fourier symbol of the normalised MoG-$3$ correlation in Eq.~\eqref{eq:mog_kernel}, and $|\widehat{\tilde p}|^{2}$ the per-mode pressure variance, normalised so $\sum_{\boldsymbol{k}}|\widehat{\tilde p}|^{2}=\sigma_p^{2}$. The two terms encode the bias-variance trade-off: $B$ kills signal modes for which the kernel-resolved gradient power $|\boldsymbol{k}|^{2}\sigma_p^{2}\widehat{\mathcal{K}}$ falls below the noise variance $\sigma_\varepsilon^{2}$, and $W$ selects the band of modes through which gradient noise leaks into the reconstruction. The domain size $L$ enters through the discrete mode set, the resolution $N$ through the upper cutoff $k_{\max}=\pi N/L$, the prior amplitude through $\sigma_p^{2}$, and the kernel shape through the normalised $\widehat{\mathcal{K}}$.

\begin{figure}[h]
    \centering
    \includegraphics[width=\textwidth]{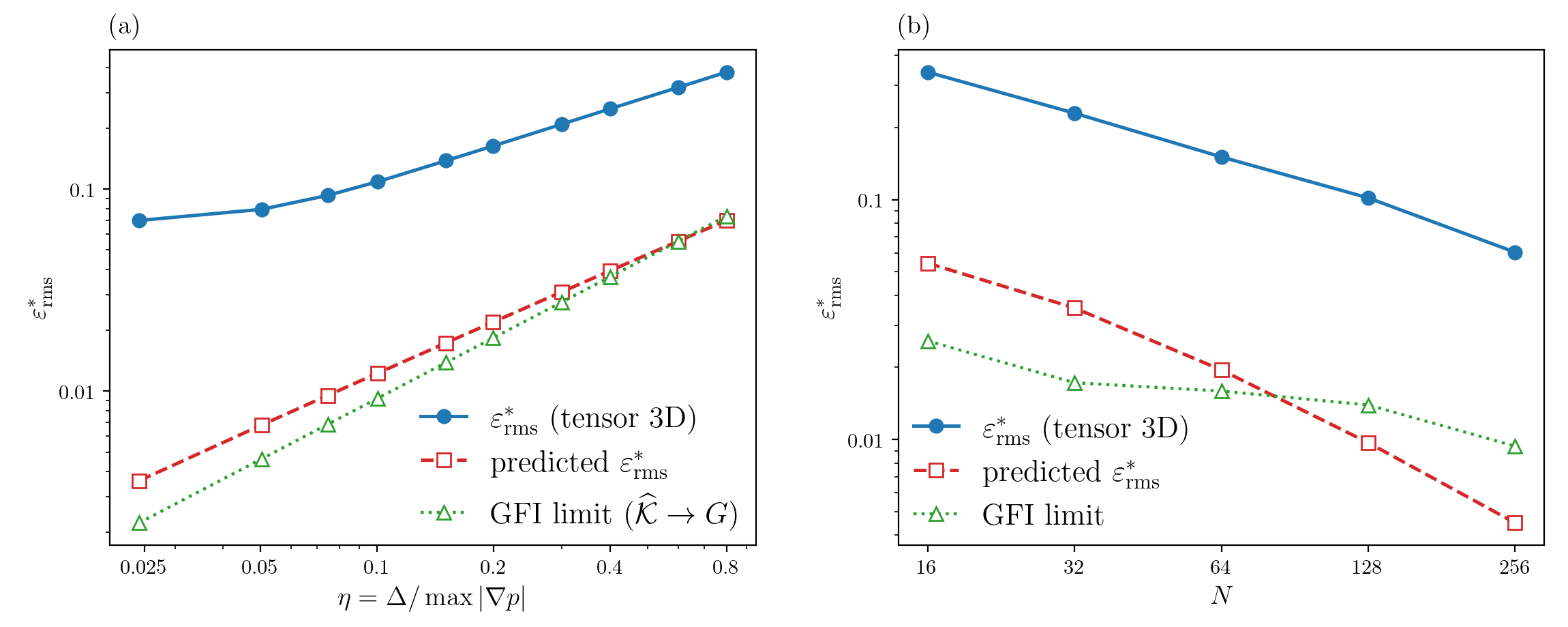}
    \caption{Trend of reconstruction error with noise and density of observation. (a) Relative RMSE $\varepsilon_{\mathrm{rms}}^{\ast}=\min_{\sigma_\varepsilon}\varepsilon_{\mathrm{rms}}$ vs gradient noise $\eta=\Delta/\max|\nabla p|$ at fixed $(N=64,L=\pi/2)$. (b) $\varepsilon_{\mathrm{rms}}^{\ast}$ vs resolution $N$ at fixed $(L=\pi/2,\eta=0.10)$. Blue circles: tensor-product 3D GPR reconstruction. Red squares: prediction from Eq.~\eqref{eq:bias_var} with $|\widehat{\tilde p}|^{2}$ taken to be the analytic empirical-kernel power spectrum $\widehat{\mathcal{K}}/L^3$, where the empirical MoG-$3$ kernel is used across all $N$. Green triangles: GFI noiseless-kernel limit.}
    \label{fig:bias_variance_prediction}
\end{figure}

Figure~\ref{fig:bias_variance_prediction}(a) validates Eq.~\eqref{eq:bias_var} on a $64^{3}$ cube of side length $L=\pi/2$ in JHTDB. The prediction (red) is computed with $|\widehat{\tilde p}|^{2}=\widehat{\mathcal{K}}/L^3$ with the fitted covariance. Both prediction and the GFI noiseless-kernel limit (green) grow nearly linearly with $\eta$ across the sweep, with the prediction sitting modestly above the GFI limit because Eq.~\eqref{eq:bias_var} also pays a kernel-bandlimit bias term that the GFI limit ignores. The measurement (blue) sits roughly constant $3\!-\!5\times$ above the prediction throughout. The gap is the boundary-contamination penalty of running the actual GPR on a non-periodic finite domain.
Panel (b) tests the resolution scaling directly. At fixed $L=\pi/2$ and $\eta=0.10$, the measurement drops monotonically with $N$, with a $5\times$ reduction across the resolution range $N\in\{16,32,64,128,256\}$. The theoretical prediction for GPR on a periodic cube shows a similar trend with uniformly lower values. The GFI limit (green) shrinks more slowly than the GPR prediction.
In conclusion, we can use Eq.~\eqref{eq:bias_var} as a rough closed-form, $\mathcal{O}(N^{d}\log N)$ lower bound that captures the noise-amplitude scaling and the resolution scaling; the absolute factor between this lower bound and the measured tensor-3D GPR is the boundary-contamination penalty for working on a non-periodic finite domain.

\section{Conclusion and future work}
\label{sec:conclusion}
We have presented a Gaussian Process Regression (GPR) framework for reconstructing pressure fields from error-embedded pressure gradient measurements. The formulation naturally avoids the need for explicit boundary conditions required by Poisson-based solvers and incorporates measurement noise through a probabilistic prior, providing both denoised reconstructions and pointwise uncertainty estimates.

Systematic comparison with the Green's Function Integration (GFI) method on forced homogeneous isotropic turbulence from the Johns Hopkins Turbulence Database demonstrates that GPR achieves accuracy comparable to GFI for well-resolved, low error level scenarios. 
{
When the spatial resolution is coarse or the error level is high, GPR with appropriate assumptions on the pressure statistics can achieve a smaller reconstruction error, although it may over-attenuate the pressure information in the intermediate- to high-wavenumber range of the energy spectrum.}
% is more accurate when the resolution is coarse, or the error level is high. 
Mathematically, GPR can be regarded as a generalization of GFI, and exhibits more systematic denoising, which suppresses measurement noise more effectively but also attenuates legitimate high-wavenumber pressure content. The impulse-response and spectral analyses attribute this behavior to the GPR operator's faster decay compared with the algebraic $m^{-1/2}$ tail of GFI, providing a built-in low-pass cut regardless of the grid resolution. The pointwise posterior standard deviation derived from the GPR posterior covariance is shown to act as a conservative envelope of the actual reconstruction error, with standardized residuals satisfying $|z|<2$ over $95\%$ of the grid on the 2D benchmark.
The framework has been extended to three dimensions via plane-wise GPR reconstructions combined with a least-squares optimization of integration constants to enforce global consistency, and in parallel via a direct tensor-product formulation whose Kronecker-structured Gram is solved by conjugate gradients.
The tensor-product GPR also outperforms GFI at coarse resolution or large error levels. A closed-form periodic-cube Wiener formula derived from the GPR operator symbol provides an $\mathcal{O}(N^d\log N)$ lower bound that captures the observed noise-amplitude and resolution scaling of the measured tensor-3D reconstruction. Beyond accuracy, the Kronecker-structured solve makes GPR affordable in 3D, only modestly more expensive than an FFT Poisson solver while retaining the innate denoising and uncertainty-quantification capabilities.

Several directions for future work remain: (a) accelerating the tensor-product matvec via the fast Gauss transform and GPU parallelization; (b) extending the matrix-free tensor-product solver of \S\ref{sec:tensor_gpr} to non-separable and anisotropic kernels; (c) investigating alternative kernel functions and mixed priors that better preserve high-wavenumber content while retaining the noise-power-gain advantage of the empirical-correlation prior.

\section*{Data and code availability}
\label{sec:data_and_code}
The GPR solvers are implemented in Python and are available on GitHub. The test data are taken from the Johns Hopkins Turbulence Database \texttt{isotropic1024coarse}.
% acknowledgment
\section*{Acknowledgments}
The support from San Diego State University is gratefully acknowledged. This project is supported by NSF-2332057.

\section*{Statement on the usage of AI tools}
During the preparation of this work, the authors used ChatGPT in order to adjust the writing style and wording accuracy of the manuscript and to align mathematical notations throughout the manuscript. After using this tool/service, the authors reviewed and edited the content as needed and take full responsibility for the content of the published article.

\appendix
\section{Equivalence between GFI and Dirichlet-energy minimization}
\label{app:gfi_dirichlet}

GFI is the Euler--Lagrange solution of a constrained Dirichlet-energy minimization. With the observed gradient field $\nabla p_{\mathrm{obs}}(\boldsymbol{x})$ on a domain $\Omega$,
\begin{equation}
    p_{\mathrm{GFI}}
    \;=\; \arg\min_{p}\tfrac{1}{2}\!\int_{\Omega}|\nabla p|^{2}\,d\boldsymbol{x},
    \qquad
    \text{s.t.}\quad \nabla p = \nabla p_{\mathrm{obs}}.
    \label{eq:dirichlet_min_problem}
\end{equation}
Introducing a vector-valued Lagrange multiplier $\boldsymbol{\mu}(\boldsymbol{x})$, the augmented functional
\begin{equation}
    \mathcal{L}[p,\boldsymbol{\mu}]
    \;=\; \tfrac{1}{2}\!\int_{\Omega}|\nabla p|^{2}\,d\boldsymbol{x}
    \;+\; \int_{\Omega}\boldsymbol{\mu}\cdot\!\bigl(\nabla p-\nabla p_{\mathrm{obs}}\bigr)\,d\boldsymbol{x}
    \label{eq:lagrangian}
\end{equation}
has stationarity conditions $-\Delta p = \nabla\!\cdot\!\boldsymbol{\mu}$ (varying $p$, assuming boundary terms vanish) and $\nabla p = \nabla p_{\mathrm{obs}}$ (varying $\boldsymbol{\mu}$). Taking the divergence of the constraint yields the Poisson equation $-\Delta p = -\nabla\!\cdot\!\nabla p_{\mathrm{obs}}$, whose solution in terms of the Green's function $G(\boldsymbol{x},\boldsymbol{x}')$ of $-\Delta$ on $\Omega$ is
\begin{equation}
    p_{\mathrm{GFI}}(\boldsymbol{x})
    \;=\; \int_{\Omega}\nabla_{\boldsymbol{x}'}G(\boldsymbol{x},\boldsymbol{x}')\cdot\nabla p_{\mathrm{obs}}(\boldsymbol{x}')\,d\boldsymbol{x}'
    \;+\; \text{boundary terms},
    \label{eq:gfi_final}
\end{equation}
which is exactly the GFI formula \eqref{eq:GFI}. Among all pressure fields consistent with the observed gradient, GFI therefore selects the one of minimum Dirichlet energy.

When the gradient data are incompatible or error-embedded, the hard constraint is relaxed to a Tikhonov form
\begin{equation}
    p_{\alpha}
    \;=\; \arg\min_{p}\tfrac{1}{2}\!\int_{\Omega}|\nabla p|^{2}\,d\boldsymbol{x}
    \;+\; \tfrac{\alpha}{2}\!\int_{\Omega}|\nabla p-\nabla p_{\mathrm{obs}}|^{2}\,d\boldsymbol{x},
    \label{eq:regularized_problem}
\end{equation}
with $\alpha>0$ trading smoothness against gradient fidelity. This is the deterministic counterpart of the GPR posterior mean: the covariance kernel plays the role of a Green's-function-like regularizer and the assumed gradient-noise variance $\sigma_\varepsilon^{2}$ corresponds to $\alpha^{-1}$.

\section{Plane-wise three-dimensional reconstruction with least-squares integration constants}
\label{appx:ls_3d}

This appendix details the plane-wise procedure of \S\ref{sec:3d_recon:planewise}. Each of the three families of axis-aligned 2D GPR slices recovers pressure only up to a per-slice additive constant; the constants are fixed by least-squares matching of the three uncorrected volumes.

\begin{algorithm}[h!]
\caption{Plane-wise 3D GPR with least-squares integration constants.\label{alg:planewise_ls}}
\KwIn{discretised gradient observations $\boldsymbol{O}_x,\boldsymbol{O}_y,\boldsymbol{O}_z$ on an $N\!\times\!N\!\times\!N$ grid (each component the corresponding $\partial p/\partial x_\alpha$ at every grid point); 2D GPR solver (\S\ref{sec:formulation}); hyperparameters $\sigma_p, \sigma_\varepsilon$ and kernel components}
\KwOut{reconstructed 3D pressure field $\boldsymbol p\in\mathbb{R}^{N\times N\times N}$}
\For(\tcp*[f]{$3N$ independent 2D solves}){$k=1,\dots,N$}{
  $\check p^{z}_{:,:,k} \leftarrow \mathrm{GPR2D}\bigl(\boldsymbol{O}_x(:,:,k),\,\boldsymbol{O}_y(:,:,k)\bigr)$\;
}
\For{$j=1,\dots,N$}{
  $\check p^{y}_{:,j,:} \leftarrow \mathrm{GPR2D}\bigl(\boldsymbol{O}_x(:,j,:),\,\boldsymbol{O}_z(:,j,:)\bigr)$\;
}
\For{$i=1,\dots,N$}{
  $\check p^{x}_{i,:,:} \leftarrow \mathrm{GPR2D}\bigl(\boldsymbol{O}_y(i,:,:),\,\boldsymbol{O}_z(i,:,:)\bigr)$\;
}
Form differences $d^{xy}_{ijk}=\check p^{y}_{ijk}-\check p^{x}_{ijk}$, $d^{xz}_{ijk}=\check p^{z}_{ijk}-\check p^{x}_{ijk}$, $d^{yz}_{ijk}=\check p^{z}_{ijk}-\check p^{y}_{ijk}$\;
Solve the sparse least-squares system $\arg\min_{\boldsymbol a,\boldsymbol b,\boldsymbol c} J$ in \eqref{eq:appx_J_d} (e.g.\ via LSMR)\;
$\acute p^{x}_{ijk}\leftarrow \check p^{x}_{ijk}+a_i$,\quad $\acute p^{y}_{ijk}\leftarrow \check p^{y}_{ijk}+b_j$,\quad $\acute p^{z}_{ijk}\leftarrow \check p^{z}_{ijk}+c_k$\;
\Return $p_{ijk}\leftarrow \tfrac{1}{3}\bigl(\acute p^{x}_{ijk}+\acute p^{y}_{ijk}+\acute p^{z}_{ijk}\bigr)$ \tcp*{$\mathcal{O}(N^{7})$ overall with dense 2D solves}
\end{algorithm}
Applying the 2D GPR of \S\ref{sec:formulation} to each slice in the three orthogonal families yields three uncorrected volumes $\check p^{x},\check p^{y},\check p^{z}\!\in\mathbb{R}^{N\times N\times N}$, where $\check p^{\alpha}_{ijk}$ is the slice-reconstruction value at $(i,j,k)$ on the constant-$\alpha$ slice. Each carries an unknown additive constant per slice; collecting these into $\boldsymbol a=(a_i)$, $\boldsymbol b=(b_j)$, $\boldsymbol c=(c_k)$, the offset-corrected volumes are
\begin{equation}
    \acute p^{x}_{ijk}=\check p^{x}_{ijk}+a_i,\quad
    \acute p^{y}_{ijk}=\check p^{y}_{ijk}+b_j,\quad
    \acute p^{z}_{ijk}=\check p^{z}_{ijk}+c_k.
    \label{eq:appx_offsets}
\end{equation}
Pointwise agreement among them is enforced by minimizing
\begin{equation}
    J(\boldsymbol a,\boldsymbol b,\boldsymbol c)
    = \sum_{ijk}\Bigl[
    \bigl(d^{xy}_{ijk}+b_j-a_i\bigr)^{2}
    +\bigl(d^{xz}_{ijk}+c_k-a_i\bigr)^{2}
    +\bigl(d^{yz}_{ijk}+c_k-b_j\bigr)^{2}\Bigr],
    \label{eq:appx_J_d}
\end{equation}
with $d^{xy}_{ijk}=\check p^{y}_{ijk}-\check p^{x}_{ijk}$, $d^{xz}_{ijk}=\check p^{z}_{ijk}-\check p^{x}_{ijk}$, $d^{yz}_{ijk}=\check p^{z}_{ijk}-\check p^{y}_{ijk}$ computed once from the uncorrected volumes.

The cost \eqref{eq:appx_J_d} is invariant under the simultaneous shift $(\boldsymbol a,\boldsymbol b,\boldsymbol c)\!\mapsto\!(\boldsymbol a+s,\boldsymbol b+s,\boldsymbol c+s)$ for any constant $s$, mirroring the fact that pressure is determined only up to an additive constant; the associated normal equations therefore have a one-dimensional null space. Stacking the $3N^{3}$ residuals of \eqref{eq:appx_J_d} into a sparse design matrix of size $3N^{3}\times 3N$ acting on the unknown vector $(\boldsymbol a,\boldsymbol b,\boldsymbol c)$, we solve the resulting least-squares problem with LSMR, which returns the minimum-norm optimum and so selects a unique representative from the null space; any residual global shift is harmless because pressure is recovered only up to a constant. As an implementation convenience, we additionally subtract the per-slice mean from each $\check p^{\alpha}$ before assembling the system, which redefines the unknowns by a known per-slice offset and improves conditioning of the LSMR iteration without altering the final field.

The final 3D field is the equally weighted average
\begin{equation}
    p_{ijk}\;=\;\tfrac{1}{3}\bigl(\acute p^{x}_{ijk}+\acute p^{y}_{ijk}+\acute p^{z}_{ijk}\bigr),
    \label{eq:appx_pmean}
\end{equation}
which minimizes the pointwise residual variance under the assumption of i.i.d.\ plane-wise reconstruction errors. Algorithm~\ref{alg:planewise_ls} summarises the complete pipeline.

\bibliographystyle{elsarticle-num-names} 
\bibliography{main}

\end{document}